\documentclass[a4paper,11pt]{article}
\pdfoutput=1
\usepackage{mathtools}
\usepackage{jheppub} 
\usepackage{axodraw2}
\usepackage{amsmath}
\usepackage{breqn}
\usepackage{url}
\usepackage{bm}
\usepackage{slashed}
   
\title{Collider Constraints on $Z^\prime$
  Models for Neutral Current $B-$Anomalies} 
\author[a]{B.C. Allanach,}
\author[b]{J. M. Butterworth,} 
\author[c]{Tyler Corbett\footnote{Corresponding author.}}

\affiliation[a]{DAMTP, University of Cambridge, Wilberforce Road, Cambridge, 
  CB3 0WA, United Kingdom}
\affiliation[b]{Department of Physics \& Astronomy, University College London, Gower St, London,
  WC1E 6BT, United Kingdom} 
\affiliation[c]{The Niels Bohr International Academy, Blegdamsvej 17, University of Copenhagen, DK-2100 Copenhagen, Denmark} 

\emailAdd{B.C.Allanach@damtp.cam.ac.uk}
\emailAdd{j.butterworth@ucl.ac.uk}
\emailAdd{corbett.t.s@gmail.com}
\preprint{MCnet-19-08}
\abstract{We examine current collider constraints on some simple $Z^\prime$ models that fit neutral current $B-$anomalies, including constraints coming from measurements of Standard Model (SM) signatures at the LHC\@. The `MDM' simplified model is not constrained by the SM measurements but {\em is}\/ strongly constrained by a 139 fb$^{-1}$ 13 TeV ATLAS di-muon search. Constraints upon the `MUM' simplified model are much weaker. A combination of the current $B_s$ mixing constraint and ATLAS' $Z^\prime$ search implies $M_{Z^\prime}>1.2$ TeV in the Third Family Hypercharge Model example case. LHC SM measurements rule out a portion of the parameter space of the model for $M_{Z^\prime}<1.5$ TeV.}

\begin{document}
\maketitle 
\flushbottom

\section{Introduction \label{sec:introduction}}

Data involving the effective Lagrangian operator $b \bar s \mu^+ \mu^-$ are
currently disagreeing with Standard Model (SM) predictions. Each individual
measurement typically disagrees at the 2-3$\sigma$ level and
over many measurements, a coherent picture is emerging. 
In particular
$R_{K^{(\ast)}} \equiv BR(B \rightarrow K^{(\ast)} \mu^+ \mu^-) / BR(B
\rightarrow K^{(\ast)} e^+ e^-)$ are predicted to be 1.00 in the SM,
for lepton invariant mass squared bin $m_{ll}^2 \in [1.1,6]$ GeV$^2$. In this bin,
current LHCb measurements~\cite{Aaij:2017vbb,CERN-EP-2019-043} imply $R_K=0.846^{+0.060}_{-0.054}{}^{+0.016}_{-0.014}$ and $R_{K^{\ast}}=0.69^{+0.11}_{-0.07}\pm0.05$. The branching
ratio $B_s \rightarrow \mu^+ \mu^-$~\cite{Aaboud:2018mst,Chatrchyan:2013bka,CMS:2014xfa,Aaij:2017vad} is also measured to be lower than the SM
prediction, which should be accurate to the percent level. Angular
distributions in the $B \rightarrow K^{(\ast)} \mu^+ \mu^-$ decays have~\cite{Aaij:2013qta,Aaij:2015oid,ATLAS-CONF-2017-023,CMS-PAS-BPH-15-008} a higher level of disagreement with SM predictions~\cite{Khachatryan:2015isa,Bobeth:2017vxj}, although here
theoretical uncertainties in the SM prediction are significant. There are several other indications of disagreements between SM
predictions and measurements and
broadly speaking, the data are consistent with a beyond-the-SM (BSM)
contribution to the $b \bar s \mu^+ \mu^-$
vertex~\cite{Altmannshofer:2013foa,Altmannshofer:2014rta,Altmannshofer:2017yso,Ciuchini:2017mik,Capdevila:2017bsm,Geng:2017svp,DAmico:2017mtc,DiChiara:2017cjq}. We
call these disagreements between measurements and SM predictions the Neutral Current $B-$Anomalies (NCBAs).
Measurements
of relevant quantities 
from Belle II with different systematic uncertainties are eagerly
awaited~\cite{Albrecht:2017odf}, as are updates from the LHC experiments. 

The operators giving BSM contributions favoured by fits to the
flavour data are 
\begin{equation}
  {\mathcal L}_{bs\mu\mu} = (\overline{b_L} \gamma^\mu s_L) (C_{LL}
  \overline{\mu_L}
  \gamma_\mu \mu_L + C_{LR} \overline{\mu_R} \gamma_\mu \mu_R) + H.c., 
\end{equation}
where $C_{LL}$ and $C_{LR}$ are Wilson coefficients, with dimensions of
inverse mass squared. 
There have been several global fits of such BSM operators that explain recent
data involving $\bar b s \bar \mu
\mu$:~\cite{Alguero:2019ptt,Alok:2019ufo,Ciuchini:2019usw,Aebischer:2019mlg,Kowalska:2019ley,Arbey:2019duh}. Details
of the 
fit methodology and results vary, but they all find that a fit involving
$C_{LL} \neq 0$ and
$C_{LR} \in [-C_{LL},\ C_{LL}]$ can provide a significant improvement over a poor fit to the
SM\@. There is evidence against
sizeable BSM operators involving $b_R$ and $s_R$ in the global fits.
For definiteness, we shall use the results of the fit of Ref.~\cite{Aebischer:2019mlg}. There,
$C_{LL}\neq 0$ only provides a good fit
  to NCBA data (6.5$\sigma$
  better than the SM prediction). A vector-like coupling (i.e.\ $C_{LL}=C_{LR}$) to muons is
a 5.8$\sigma$ better fit than the SM at the best-fit point, whereas an axial
coupling ($C_{LL}=-C_{LR}$) 
coupling to muons is 5.6$\sigma$ better
than the SM at the best-fit point.

At tree-level, a BSM contribution to $C_{LL}$ or $C_{LR}$ can come
from leptoquarks and/or $Z^\prime$s, either of which must have flavour dependent couplings. Here, we
shall focus on the $Z^\prime$ possibility. Many models based on spontaneously
broken flavour-dependent gauged $U(1)$ symmetries~\cite{Ellis:2017nrp,Allanach:2018vjg} have been
proposed 
from which such $Z^\prime$s may result, for example from $L_\mu-L_\tau$ and
related groups~\cite{Gauld:2013qba,Buras:2013dea,Buras:2013qja,Altmannshofer:2014cfa,Buras:2014yna,Crivellin:2015mga,Crivellin:2015lwa,Sierra:2015fma,Crivellin:2015era,Celis:2015ara,Greljo:2015mma,Altmannshofer:2015mqa,Allanach:2015gkd,Falkowski:2015zwa,Chiang:2016qov,Becirevic:2016zri,Boucenna:2016wpr,Boucenna:2016qad,Ko:2017lzd,Alonso:2017bff,Alonso:2017uky,1674-1137-42-3-033104,Bonilla:2017lsq,Bhatia:2017tgo,Ellis:2017nrp,CHEN2018420,Faisel:2017glo,PhysRevD.97.115003,Bian:2017xzg,PhysRevD.97.075035,King:2018fcg,Duan:2018akc,Allanach:2018lvl,Allanach:2018odd}. Some
models also have several abelian groups~\cite{Crivellin:2016ejn} leading to
multiple $Z^\prime$s. Some other models~\cite{Kamenik:2017tnu,Camargo-Molina:2018cwu} generate the $b
\bar s \mu^+ \mu^-$ operator with a loop-level penguin diagram. 

In Ref.~\cite{Chivukula:2017qsi}, Run I di-jet and di-lepton resonance 
searches (and early Run II searches) were used to
constrain simple $Z^\prime$ models that fit the NCBAs.
In Refs.~\cite{Allanach:2017bta,Allanach:2018odd}, the sensitivity of future
hadron colliders to $Z^\prime$ models that fit the NCBAs was estimated. A 100
TeV future circular collider
(FCC)~\cite{Mangano:2018mur} would have sensitivity to the whole of  
parameter space for one model (MDM)
and the majority of parameter space for another (MUM).
However, given recent updates on
LHC $Z^\prime$ searches released by the ATLAS experiment and
on the NCBAs,
it seems that the time is ripe for a fresh
analysis of the resulting constraints upon $Z^\prime$ models that fit the
NCBAs. 

ATLAS has  released 13 TeV
36.1~fb$^{-1}$  $Z^\prime \rightarrow t \bar t$
searches~\cite{Aaboud:2018mjh,Aaboud:2019roo}, which impose $\sigma \times BR(Z^\prime \rightarrow t \bar t)<10$
fb for large $M_{Z^\prime}$. There is also a search~\cite{Aad:2015osa} for
$Z^\prime 
\rightarrow \tau^+ \tau^-$ for 10 fb$^{-1}$ of 8 TeV data, which rules out
$\sigma \times BR(Z^\prime \rightarrow \tau^+ \tau^-)<3$ fb for large
$M_{Z^\prime}$. These searches constrain, in principle, some of the flavourful
$Z^\prime$ models that we introduce below, but they produce less stringent
constraints upon the models that we study than an ATLAS search for $Z^\prime
\rightarrow \mu^+\mu^-$ in 139
fb$^{-1}$ of 13 TeV $pp$ 
collisions~\cite{Aad:2019fac}. We shall therefore concentrate upon this
search, recasting it for some models that solve the NCBAs. The constraints are
in the form of upper limits
upon the fiducial cross-section $\sigma$ times branching ratio
to di-muons $BR(Z^\prime \rightarrow \mu^+ \mu^-)$
as a function of $M_{Z^\prime}$. 
At large $M_{Z^\prime} \approx 6$ TeV, $\sigma \times BR(Z^\prime
\rightarrow \mu^+\mu^-)<0.015$ fb~\cite{atlasData} and indeed this will prove to be the most stringent
$Z^\prime$ direct search constraint (being stronger than the others mentioned
above) on the models which we study.

In~\S~\ref{sec:simp}, we introduce simplified models $Z^\prime$ which can
provide a good fit to the NCBAs, examining the important $B_s$ mixing
constraint in \S~\ref{sec:constraints}. In \S~\ref{sec:def}, we define the
mixed-up muon (MUM) and mixed-down muon (MDM) simplified models, followed by
the more complete Third Family Hypercharge Model (TFHM). In \S~\ref{sec:meth}, we
describe how we recast the ATLAS $Z^\prime \rightarrow \mu^+ \mu^-$ search and
outline how other Run I and Run II measurements are checked against the model.
Example parameter space points for each model are listed for illustration in
\S~\ref{sec:results}, before the combined collider constraints upon the models
are presented. We summarise in \S~\ref{sec:conc}. In Appendix~\ref{sec:fdef}
we define the fields. Properties of the three models studied throughout their
parameter space are relegated to Appendix~\ref{sec:propMUM}.

\section{Models and Constraints\label{sec:simp}}

We consider two representative models of $Z^\prime$s,
following Ref.~\cite{Allanach:2017bta}, which introduced the na\"ive and the $33\mu\mu$
models. The tree-level $Z^\prime$ Lagrangian couplings that should be present
in $Z^\prime$ models 
in order to explain the NCBAs are
\begin{equation}
\mathcal{L}_{Z^\prime f} =\left( g_{sb} \overline{s_L} \slashed{Z}^\prime b_L  +
  \text{h.c.} \right)  + g_{\mu\mu} \overline{\mu_L}\slashed{Z}^\prime \mu_L + 
\ldots \label{wrongNaive}
  \end{equation}
A global fit to NCBAs and $V_{ts}$ in
Ref.~\cite{Aebischer:2019mlg} found that   
the couplings and masses of $Z^\prime$ particles are constrained to be
\begin{equation}
g_{sb}g_{\mu\mu} = -x \left(\frac{M_{Z^\prime}}{36\text{TeV}}
\right)^2,
\label{constraint}
  \end{equation}
if $g_{sb}$ and $g_{\mu\mu}$ are real, where $x=1.06 \pm 0.16$ in the recent
fit to the NCBAs from Ref.~\cite{Aebischer:2019mlg}. 
Throughout this paper, we shall enforce Eq.~\ref{constraint}, typically taking the 
central value from the fit. 
In general, $g_{sb}$ and $g_{\mu\mu}$ are complex. However, here, we take
$g_{\mu\mu}$ to be real and positive and $g_{sb}$ to be negative.
In the models we introduce below, $g_{sb}$
may have a 
small imaginary part. Since the full effects of complex phases
are outside the scope of this work, whenever we refer to $g_{sb}$ below, we
shall implicitly refer to the real part of its value.

\subsection{$B_s$ mixing constraint \label{sec:constraints}}

\begin{figure}
\center
\includegraphics[scale=.15]{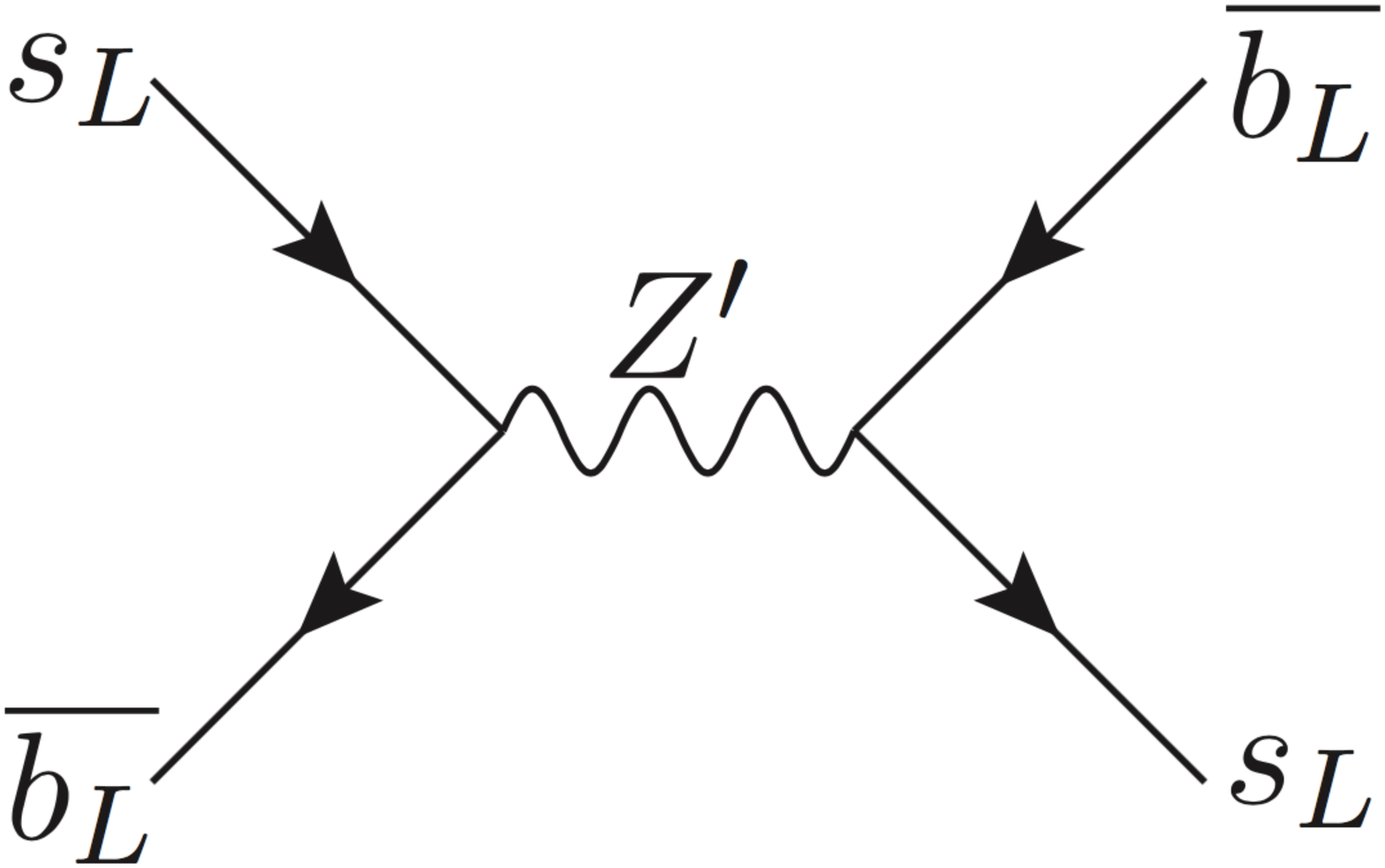}
\caption{Feynman diagram of the tree-level $Z^\prime$ contribution to $B_s-\overline{B_s}$
  mixing. \label{fig:dms}}  
\end{figure}
$Z^\prime$ models are subject to a number of constraints,
a particularly strong one
originating from measurements of  $B_s-\overline{B_s}$ mixing, which
constrains a function of $g_{sb}$ and $M_{Z^\prime}$.
A Feynman diagram depicting the $Z^\prime$ contribution is shown in
Fig.~\ref{fig:dms}. 
The bound on a non-SM contribution depends upon the
hadronic decay constant $f_{B_s}$ and bag parameter $B_s$.
The experimental measurement of the mixing parameter $\Delta M_s$ is~\cite{Amhis:2016xyh}
$\Delta M_s^{exp} = (17.757 \pm 0.021) \text{~ps}^{-1}$.
We use a determination of the SM prediction using recent lattice  data
and sum rules~\cite{King:2019lal}
\begin{equation}\Delta M_s^{SM} = (18.5^{+1.2}_{-1.5}) \text{~ps}^{-1}. \label{dmssm}\end{equation}
In order to calculate the resulting bound on $Z^\prime$ models, we follow
Ref.~\cite{DiLuzio:2017fdq}. 
In a model inducing the BSM operator
\begin{equation}
{\mathcal L}^{NP} = \frac{-4 G_F}{\sqrt{2}} (V_{tb} V_{ts}^\ast)^2
\left[ c_{sb}^{LL} (\overline{s_L} \gamma_\mu b_L)(\overline{s_L} \gamma^\mu
  b_L)+H.c.\right],
\end{equation}
where $G_F=1.1663787(6) \times 10^{-5}$ GeV$^{-2}$ is the Fermi coupling
constant, the SM prediction of $B_s$ mixing is modified to
$\Delta M_s^{pred}=|1 + c_{sb}^{LL} / R^{loop}_{SM} | \Delta M_S^{SM}$,
where
$R^{loop}_{SM}=1.3397 \times 10^{-3}$.
Our flavour changing $Z^\prime$s induce the Wilson coefficient 
\begin{equation}
  c_{sb}^{LL} = \frac{\eta^{LL}}{4 \sqrt{2} G_F M_{Z^\prime}^2}
   \frac{g_{sb}^2}{(V_{tb}V_{ts}^\ast)^2}, \label{WC}
\end{equation}
$\eta^{LL}$ takes renormalisation between $M_{Z^\prime}$
and $M_Z$ into account. It is a slow (logarithmic) function of $M_{Z^\prime}$:
$\eta^{LL}=0.79$ for $M_{Z^\prime}=1$ TeV, whereas $\eta^{LL}=0.75$ for
$M_{Z^\prime}=10$ TeV (we shall be concerned here with $M_{Z^\prime} \leq 6$ TeV). Here, we shall take $\eta^{LL}=0.79$ whatever
$M_{Z^\prime}$, since this value gives the stronger limit out of the two
numbers quoted and since $\eta^{LL}$ is quite insensitive to $M_{Z^\prime}$  
anyway. 
Eq.~\ref{dmssm} implies the 2$\sigma$ lower bound $\Delta M_S^{SM}>15.5$
ps$^{-1}$, 
leaving room for a BSM contribution to make up a shortfall to the
experimental $2\sigma$ upper bound if (by substituting
$|V_{ts}^\ast V_{tb}|=0.04$ into Eq.~\ref{WC})\footnote{This inferred bound has changed in recent years due to changes in data
and lattice inputs: pre-2016, the denominator was $148$
TeV~\cite{DAmico:2017mtc}, whereas from 2016-2019
the inferred denominator became 600 TeV~\cite{DiLuzio:2017fdq}.}
\begin{equation}
|g_{sb}|\lesssim
M_{Z^\prime}/(194\text{~TeV}). \label{bsmix}
\end{equation} 
This places a strong constraint upon $Z^\prime$ models that explain the
NCBAs~\cite{DiLuzio:2017fdq}.

\subsection{Model definitions and couplings \label{sec:def}}
Following Ref.~\cite{Allanach:2018odd}, we begin with simplified models originating from assuming that the $Z^\prime$ 
only couples to left-handed quarks and to left-handed
leptons. Our direct search collider constraints are not strongly dependent
upon the 
spin-structure of the $Z^\prime$ couplings and so this model should suffice to
cover others (for example sharing the BSM operator between left-handed and
right-handed muons).  
The $Z^\prime$ couplings to the mass eigenstate fermions 
in the model are  
\begin{equation}
\mathcal{L} =
\overline{\bm u_L} V \Lambda^{(Q)} V^\dagger \slashed{Z}^\prime
{\bm u_L} + 
\overline{\bm d_L} \Lambda^{(Q)} \slashed{Z}^\prime {\bm d_L}
+ 
\overline{\bm \nu_L} U \Lambda^{(L)} U^\dagger \slashed{Z}^\prime {\bm \nu_L} + 
\overline{\bm e_L} \Lambda^{(L)} \slashed{Z}^\prime {\bm e_L},
\label{secSU2}
  \end{equation}
where we have written the Cabibbo-Kobayashi-Maskawa (CKM) matrix as $V$ and the
Pontecorvo-Maki-Nakagawa-Sakata matrix as $U$
(see Appendix~\ref{sec:def} for field definitions). 
$\Lambda^{(Q)}$ and $\Lambda^{(L)}$ are 3 by 3 matrices of dimensionless
couplings. In order to reproduce a $Z^\prime$ coupling to left-handed muons,
as required to fit the $B-$anomalies, we use 
\begin{equation}
\Lambda^{(L)} = g_{\mu\mu} \left( \begin{array}{ccc} 
0 & 0 & 0 \\
0 & 1 & 0 \\
0 & 0 & 0 \\ \end{array}
\right), 
\end{equation}
The two simplified models introduced involve two different limiting
assumptions for $\Lambda^{(Q)}$, in order to provide an estimate of how much
the assumption changes predictions:
\begin{enumerate}
\item
{\bf The `mixed-up-muon' (MUM) model}, with
\begin{equation}
\Lambda^{(Q)} = g_{sb} \left( \begin{array}{ccc} 
0 & 0 & 0 \\
0 & 0 & 1 \\
0 & 1 & 0 \\ \end{array}
\right),
\end{equation}
\item
{\bf The `mixed-down-muon' (MDM) model}, with
\begin{equation}
\Lambda^{(Q)} = g_{tt} V^\dagger \cdot \left( \begin{array}{ccc} 
0 & 0 & 0 \\
0 & 0 & 0 \\
0 & 0 & 1 \\ \end{array}
\right) \cdot V
\end{equation}
Matching
$\Lambda^{(Q)}$ here with Eq.~\ref{wrongNaive} identifies
\begin{equation}
g_{sb}=V_{ts}^\ast V_{tb} g_{tt}. \label{gsb}
\end{equation}
In the present article, we are not concerned with the effects of small complex
phases: we shall take $g_{tt}$ to be real\footnote{Although we include the effects
of phases in the CKM matrix in our numerical simulations, they are not
important for our results and we ignore them in analytic discussion.}.
$g_{tt}>0$ ensures $g_{sb}<0$ as required by Eq.~\ref{constraint}, since 
$V_{ts}\approx -0.04$ and $V_{tb} \approx 1$.
\end{enumerate}
We may characterise the MUM and MDM simplified models by three important
parameters: $M_{Z^\prime}$, $|g_{sb}|$ and $g_{\mu\mu}$. In practice, we shall
use $M_{Z^\prime}$ and $|g_{sb}|$, whilst fixing $g_{\mu\mu}$ so as to fit the
central values of the NCBAs in Eq.~\ref{constraint}. We note here that, since
the MUM and MDM models are simplified, in reality the $Z^\prime$ might have
more couplings than the ones introduced and so could be wider than predicted
in the strict MUM or MDM limit. One could, instead of calculating the
$Z^\prime$ width $\Gamma$, use the MUM or MDM limit as a lower bound and allow
it to vary independently of $g_{sb}$ and $g_{\mu\mu}$. We expect that
increasing $\Gamma$ will weaken search constraints, and so in some sense,
neglecting this `additional width' effect (which is the approach we shall take) is conservative. 

{\bf The Third Family Hypercharge Model (TFHM)}
is based~\cite{Allanach:2018lvl} on a $U(1)_F$
gauge extension to the Standard Model, only the Higgs doublet,
a new complex scalar SM singlet and
third family fermions have non-zero $U(1)_F$ quantum numbers.
The heavy $Z^\prime$ comes from spontaneously breaking the $U(1)_F$ and it is thus
a more complete model than the MUM and MDM models. The model explains, in
broad brush-strokes, the hierarchical heaviness of the third family of charged
fermions and the smallness of CKM mixing angles. 
Anomaly
cancellation implies that the $U(1)_F$ quantum numbers of the third
family fields are proportional to their hypercharges.
The $Z^\prime$ couplings are, up to corrections $\mathcal{O}
\left({M_Z^2}/{{M_Z^\prime}^2}\right)$
\begin{eqnarray}
\mathcal{L}_{X \psi} =&g_F&\left( 
\frac{1}{6}\overline{\bm u_L} \Lambda^{(u_L)} \slashed{Z}^\prime {\bm u_L} + 
\frac{1}{6}\overline{\bm d_L} \Lambda^{(d_L)} \slashed{Z}^\prime {\bm d_L}-\frac{1}{2}
\overline{\bm \nu_L} \Lambda^{(\nu_L)} \slashed{Z}^\prime {\bm \nu_L} -\frac{1}{2}
\overline{\bm e_L} \Lambda^{(e_L)} \slashed{Z}^\prime {\bm e_L}\right. \nonumber \\
&&\left. +\frac{2}{3}
\overline{\bm u_R} \Lambda^{(u_R)} \slashed{Z}^\prime {\bm u_R}-\frac{1}{3}
\overline{\bm d_R} \Lambda^{(d_R)} \slashed{Z}^\prime {\bm d_R}-
\overline{\bm e_R} \Lambda^{(e_R)} \slashed{Z}^\prime {\bm e_R}
\right),  
  \end{eqnarray}
where we have 
defined the 3 by 3 dimensionless Hermitian coupling matrices 
\begin{equation}
\Lambda^{(I)} \equiv V_{I}^\dagger \xi V_{I} ,
\label{lambdas}
\end{equation}
$I \in \{ u_L, d_L, e_L, \nu_L, u_R, d_R, e_R \}$ and
\begin{equation}
\xi = \left(\begin{array}{ccc}
0 & 0 & 0 \\ 0 & 0 & 0 \\ 0 & 0 & 1 \\
\end{array}\right).
\end{equation}
The $V_I$ are unitary 3 by 3 matrices in family space and $g_F$ is the
dimensionless gauge
coupling of $U(1)_F$.
For definiteness, we shall
examine the phenomenological example case introduced in
Ref.~\cite{Allanach:2018lvl}: 
\begin{equation}
V_{d_L}=\left( \begin{array}{ccc}
1 & 0 & 0 \\
0 & \cos \theta_{sb} & -\sin \theta_{sb} \\
0 & \sin \theta_{sb} & \cos \theta_{sb} \\
\end{array}\right)
\qquad\text{and}\qquad
V_{e_L}=\left( \begin{array}{ccc}
1 & 0 & 0 \\
0 & 0 & 1 \\
0 & 1 & 0 \\
\end{array}\right),
\label{vdl}
\end{equation}
$V_{u_L}= V_{d_L}V^\dagger$, $V_{u_R}=V_{d_R}=V_{e_R}=1$.
To summarise, in the TFHM example case (TFHMeg), we have free parameters $|g_F|$,
$M_{Z^\prime}$ and $\theta_{sb}$. In practice, we vary $M_{Z^\prime}$ and
$\theta_{sb}$, setting $g_F$ so as to satisfy the central value of the NCBAs,
i.e.\ Eq.~\ref{constraint}, which translates to
\begin{equation}
  g_F = \frac{M_{Z^\prime}}{36 \text{~TeV}} \sqrt{\frac{24 x}{\sin(2
      \theta_{sb})}} \label{y3const}
\end{equation}
with $x=1.06$.

\section{Re-casting Collider Constraints \label{sec:meth}}

In its recent $Z^\prime \rightarrow \mu^+ \mu^-$ search, ATLAS defines~\cite{Aad:2019fac} a
fiducial cross-section $\sigma$ where each muon has transverse momentum $p_T>30$ GeV
and pseudo-rapidity $|\eta|<2.5$. The di-muon invariant mass, $m_{\mu \mu}>225$
GeV. No evidence for a significant bump in $m_{\mu\mu}$ was found, and so
95$\%$ upper limits on $\sigma \times BR(\mu^+ \mu^-)$.

\begin{figure}
  \begin{center}
    \includegraphics[width=12cm]{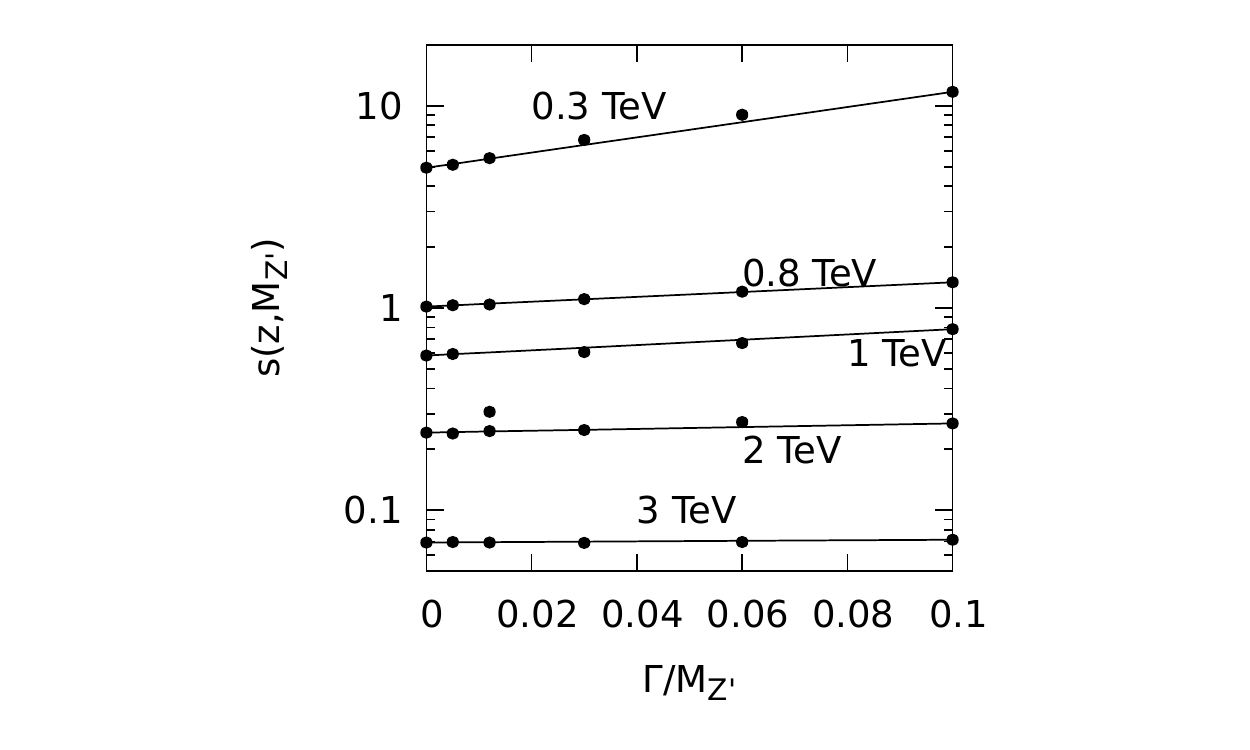}
  \end{center}
  \caption{\label{fig:fit} Examples of the fit in Eq.~\ref{lims} (shown by
    lines) compared to
    ATLAS data (shown by points), for various values of $M_{Z^\prime}$, shown
    as a label by each line.}
  \end{figure}
Re-casting constraints from such a bump-hunt in different $Z^\prime$ models is
fairly simple: we must just calculate $\sigma \times BR(\mu^+ \mu^-)$ for the
model in question and apply the bound at the relevant value of $M_{Z^\prime}$
and $\Gamma/M_{Z^\prime}$. Efficiencies are taken into account in the
experimental bound and so there is no need for us to perform a detector
simulation. 
For generic $z \equiv\Gamma/M_{Z^\prime}$, we
interpolate/extrapolate the upper bound $s(z, M_{Z^\prime})$ on $\sigma \times BR(\mu^+
\mu^-)$ from those given by ATLAS at $z=0$ and $z=0.1$. In practice, we use a linear
interpolation in $\ln s$:
\begin{equation}
  s(z,M_{Z^\prime}) = s(0,M_{Z^\prime})
  \left[ \frac{s(0.1,M_{Z^\prime})}{s(0,M_{Z^\prime})}\right]^{\frac{z}{0.1}}.  \label{lims}
  \end{equation}
Fig.~\ref{fig:fit} shows examples of such a fit for five different values of
$M_{Z^\prime}$ compared to ATLAS upper limits. One point not lying on the line
is due to a statistical fluctuation in data, but generally, the figure validates
Eq.~\ref{lims} as being a reasonable fit within the range
$\Gamma/M_{Z^\prime} \in [0,0.1]$. In general, we shall also use Eq.~\ref{lims} to
extrapolate out of this range, however this will only turn out to play a
r\^{o}le in part of the TFHMeg parameter space, which we shall delineate. 

\begin{figure}
\centering
\includegraphics[scale=.15]{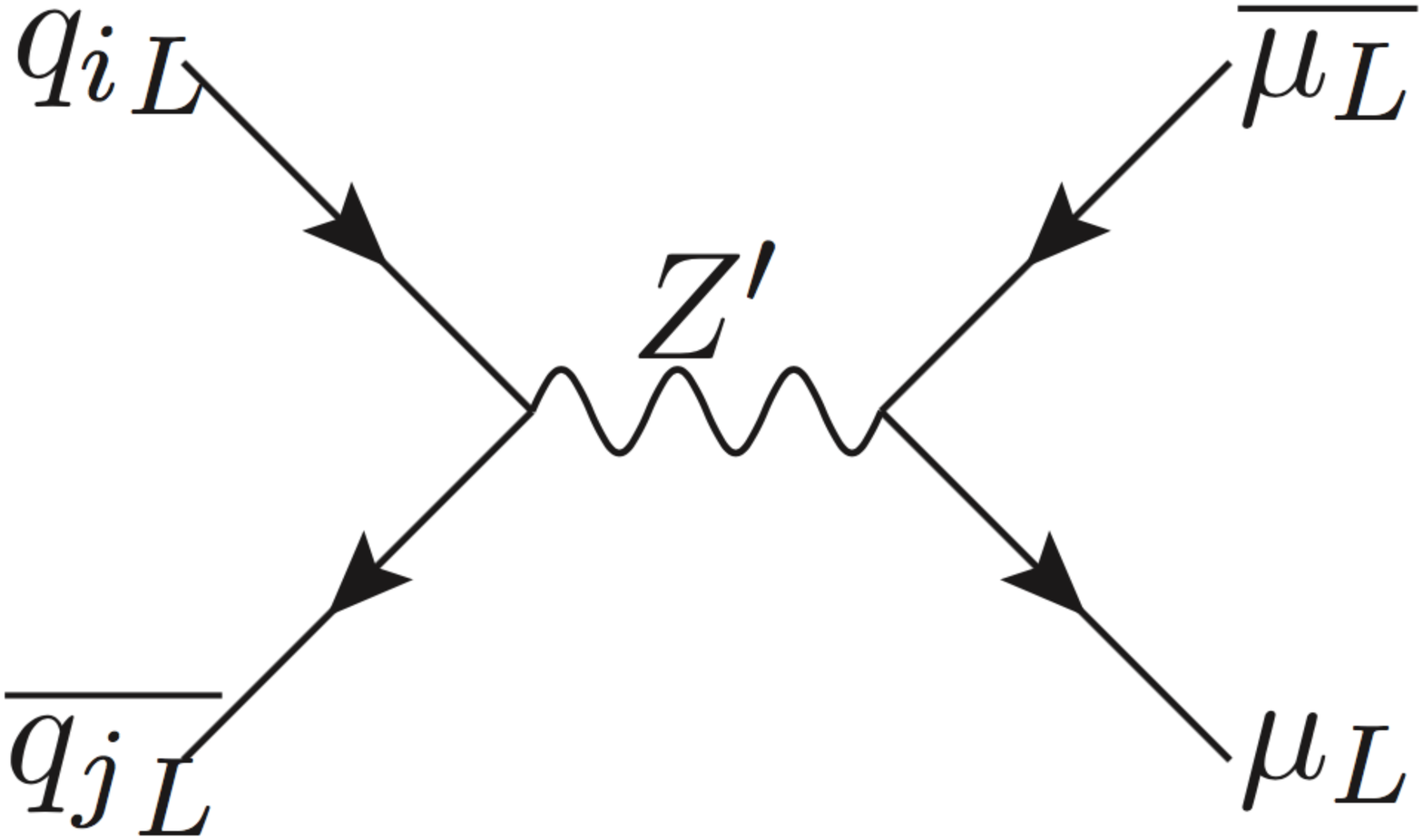}
\caption{Feynman diagram of tree-level $Z^\prime$ production in the
  LHC, where $q_{i,j} \in \{ u,c,d,s,b \}$ are such that the
  combination ${q_i}_L \overline{{q_j}_L}$  has zero electric
  charge.\label{fig:prod}}    
\end{figure}

For the TFHMeg, we made a {\tt UFO} file\footnote{The {\tt UFO} file is 
  included in the ancillary information submitted with the {\tt arXiv} version
of this paper.} by using
\texttt{FeynRules}~\cite{Degrande:2011ua,Alloul:2013bka}.
The MUM model and MDM model files are taken from
Ref.~\cite{Allanach:2018odd}. These {\tt UFO} files allow the {\tt MadGraph}
calculation of $\sigma \times BR(Z^\prime \rightarrow \mu^+ \mu^-)$ by
\texttt{MadGraph_2_6_5}~\cite{Alwall:2014hca}.
{\tt MadGraph}
estimates $\sigma \times BR(Z^\prime \rightarrow \mu^+ \mu^-)$ of
the
tree-level production processes shown in Fig.~\ref{fig:prod} in 13 TeV centre of mass energy
$pp$ collisions.
We use 5-flavour parton distribution
functions 
in order to re-sum the logarithms associated with the initial state
$b$-quark~\cite{Lim:2016wjo}. 

\subsection{Constraints from {\sc Contur}}

Introducing the BSM terms discussed above leads to other possible new processes and signatures in $pp$ collisions in addition to the
di-muon channel already considered.
For example, in the TFHMeg model, the $Z^\prime$ has a branching fraction in the range 10-20\% to $b\bar{b}$, up to 40\% to $t\bar{t}$ 
and 20-30\% to $\tau^+\tau^-$. It is often produced in association with additional $b$-jets, and the cross section for associated 
production with an isolated photon can be as high as a few femtobarns. Many relevant measurements of such signatures have already 
been made by the LHC experiments, and we use the {\sc Contur}~\cite{Butterworth:2016sqg} 
tool to check whether these measurements already disfavour any of the parameter space of our model.
We use {\tt Herwig7}~\cite{Bellm:2015jjp,Bellm:2017bvx} and its {\tt UFO} interface to calculate the cross section for all the new processes implied
at the LHC by our models, and to inclusively generate the implied events. These events are then passed to Rivet~\cite{Buckley:2010ar} 
version 2.7, which contains an extensive library of particle-level collider measurements, especially from LHC Run I but also increasingly 
now from Run II\@. While these will not be as sensitive as individual searches using the full data set, they have the advantage of relative
model-independence and ease of reinterpretation. All these measurements are in agreement with the SM, and {\sc Contur} therefore
treats them as SM background to a potential contribution from our models, evaluating whether the presence of an additional BSM contribution
(in particular a $Z^\prime$ mass peak) would have been visible within the experimental uncertainty. This is then converted into an
exclusion limit. Previous studies~\cite{Butterworth:2016sqg,Brooijmans:2018xbu,Amrith:2018yfb} have shown 
that this approach typically gives a comparable sensitivity to dedicated searches and can sometimes 
pick up unexpected additional signatures.

\section{Results \label{sec:results}}

\begin{table}
  \begin{center}
    \begin{tabular}{|c|c||c|c||c|c|}\hline
      \multicolumn{2}{|c||}{MDM} &
      \multicolumn{2}{|c||}{MUM} &
      \multicolumn{2}{|c|}{TFHMeg} \\ \hline \hline
      parameter & value & parameter & value & parameter & value \\ \hline
      $M_Z^\prime$ & 1 TeV & $M_{Z^\prime}$ & 0.8 TeV & $M_{Z^\prime}$ & 1.9
      TeV \\
      $g_{sb}$ & -0.001 & $g_{sb}$ & 0.01 & $\theta_{sb}$ & 0.08 \\ 
      $g_{\mu\mu}$ &0.82 & $g_{\mu\mu}$ & -0.052 & $g_F$ & 0.67 \\ \hline
      quantity & value & quantity & value & quantity & value \\ \hline
      $\Gamma/M_{Z^\prime}$ &0.018 & $\Gamma/M_{Z^\prime}$ & 8.4$\times
      10^{-5}$ & $\Gamma/M_{Z^\prime}$ & 0.020\\
      $BR( \mu^+ \mu^-)$ & 0.50 & $BR(\mu^+ \mu^-)$ &0.43 &$BR(\mu^+ \mu^-)$ &
      0.08\\
            $BR( t \bar t)$ & 1.3$\times 10^{-3}$ & $BR(t \bar t)$ &2.6$\times 10^{-4}$ & $BR(t \bar t)$ & 0.42\\
            $BR( b \bar b)$ & 1.3$\times 10^{-3}$ & $BR(b \bar s+s \bar b)$
      &0.09 &$BR(b \bar b)$ & 0.13\\
             & & $BR(t \bar c+c \bar t)$ &0.04 &$BR(\tau^+ \tau^-)$ & 0.30\\ \hline
channel & $\sigma \times BR$/fb &channel & $\sigma \times BR$/fb &
channel & $\sigma \times BR$/fb \\ \hline
$b \bar b$ &0.31 & & &       $b \bar b$ & 0.13 \\
$\bar s b+\bar b s$ & 0.001 & $\bar s b+\bar b s$ & 0.45 &       $\bar s b+\bar b s$ & 0.001 \\
 & & $\bar c c$ & 0.002 &       &       \\
 & &  & &        &        \\ \hline
total & 0.31 &total &0.45 &        total & 0.13 \\
upper limit & 0.37 & upper limit &0.92 & upper limit & 0.25\\ \hline
    \end{tabular}
  \end{center}
  \caption{\label{tab:exampleSig}{Illustration of
      example points in parameter space. The third parameter listed in each case is derived in terms of the
      two above it by the best fit to the NCBAs, i.e.\ Eq.~\ref{constraint}
      with $x=1.06$.
      `channel' lists the contribution to the total $Z^\prime$ cross-section
      times branching ratio from the various quark parton distribution
      functions (PDFs). 
      For each production mode,
  we list the $Z^\prime$ fiducial production cross-section times branching
  ratio into muon pairs $\sigma \times BR$. Other production modes have cross-sections that are
  smaller than $10^{-3}$ fb. The upper limit on the cross-section is the
  95$\%$ $CL_s$ bound derived
  from the ATLAS di-muon search~\cite{Aad:2019fac} according to Eq.~\ref{lims}.}}
  \end{table}
In Table~\ref{tab:exampleSig}, we display one point for each model
studied, where the model parameters are chosen to fit the NCBAs and to be close
to the exclusion of the ATLAS di-muon search in each case. We see that each
point has a narrow $Z^\prime$: $\Gamma/M_{Z^\prime}\leq 0.02$ (however, there
are other points with larger values, as we shall see).
In each model, the branching ratio into neutrinos is identical to that of
muons and tagging an additional jet would result in a monojet $Z^\prime
\rightarrow$ invisible signature at the LHC\@.  In the MUM model, we note
the possible flavour changing channels $Z^\prime \rightarrow t \bar c + \bar c
t$,
$Z^\prime \rightarrow b \bar s + s \bar b$, which could also be used for
searches. In the TFHMeg, decays to top pairs are 6 times more prevalent than
those into muon pairs, which could prove to be an important channel for
searches, as could decays into tau pairs (4 times more prevalent than muon
pairs). Although these channels have a higher branching ratio than di-muons,
the current bounds are sufficiently weaker such that di-muons (the only
channel currently having been analysed for the full 139 fb$^{-1}$ LHC Run II
dataset) provide the strongest constraint.
The table is
instructive by exemplifying which PDFs are important for $Z^\prime$ production
in each case. In the MDM model and the TFHMeg, $b \bar b\rightarrow Z^\prime$ dominates,
whereas in the MUM model, $b \bar s\rightarrow Z^\prime$ dominates. The upper
limit from the ATLAS di-muon search is shown for the particular $M_{Z^\prime}$
of the parameter point, for the narrow width limit. In what follows, we
include the dependence of these upper limits upon the width, as described in \S~\ref{sec:meth}. 

\begin{figure}
\begin{center}
  \unitlength=\textwidth
  \begin{picture}(1,0.9)(0.1,0.05)
    \put(0.25,0){\includegraphics[width=0.8 \textwidth]{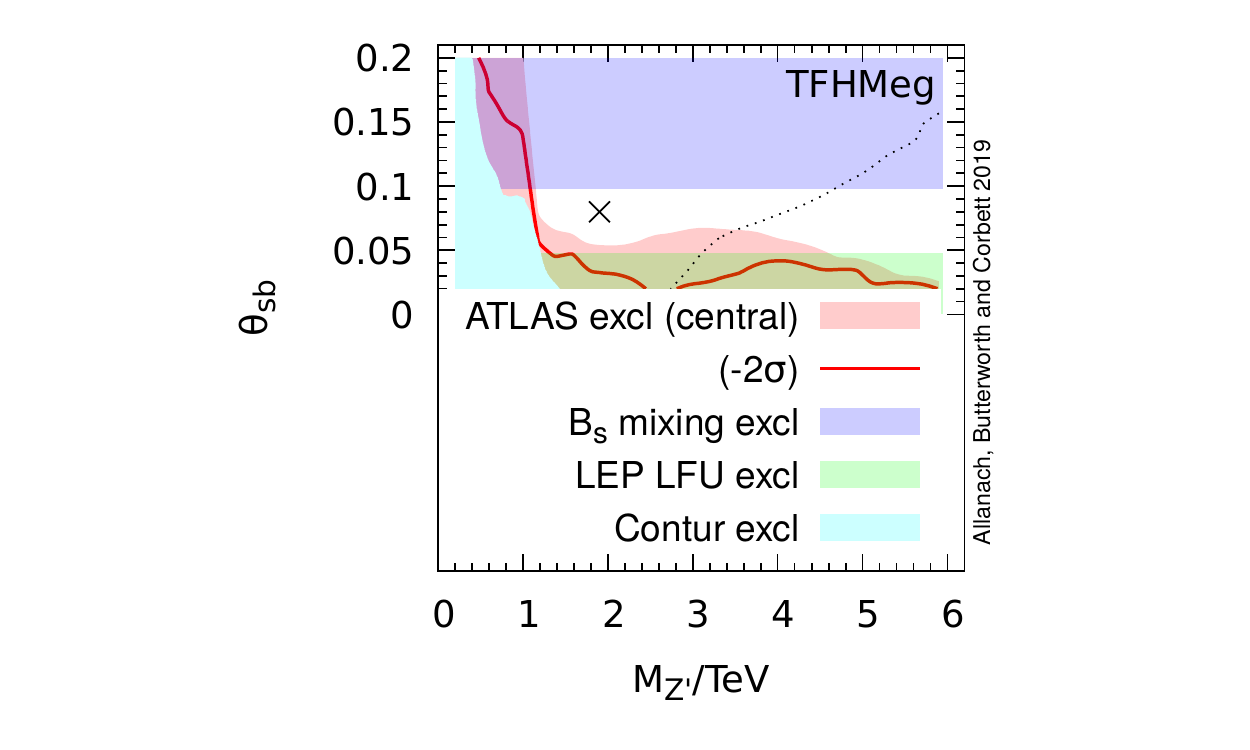}}
    \put(-0.05,0.5){\includegraphics[width=0.8 \textwidth]{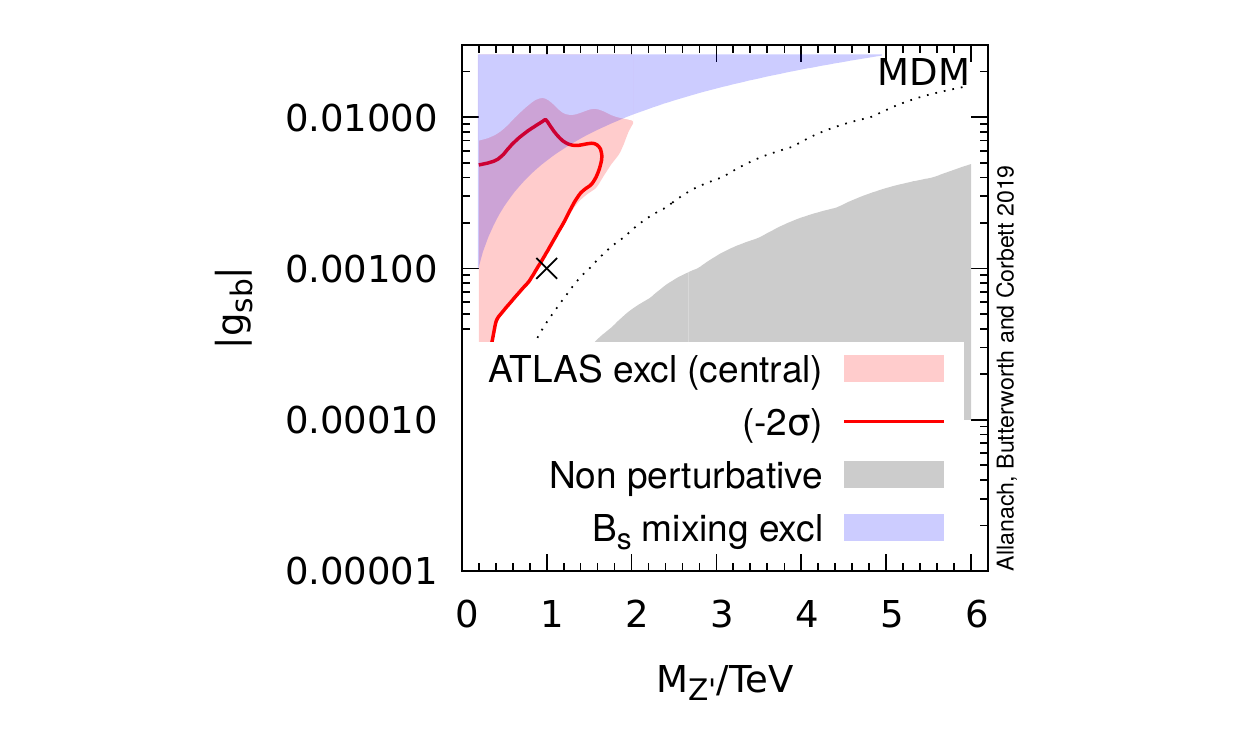}}
    \put(0.5,0.5){\includegraphics[width=0.8 \textwidth]{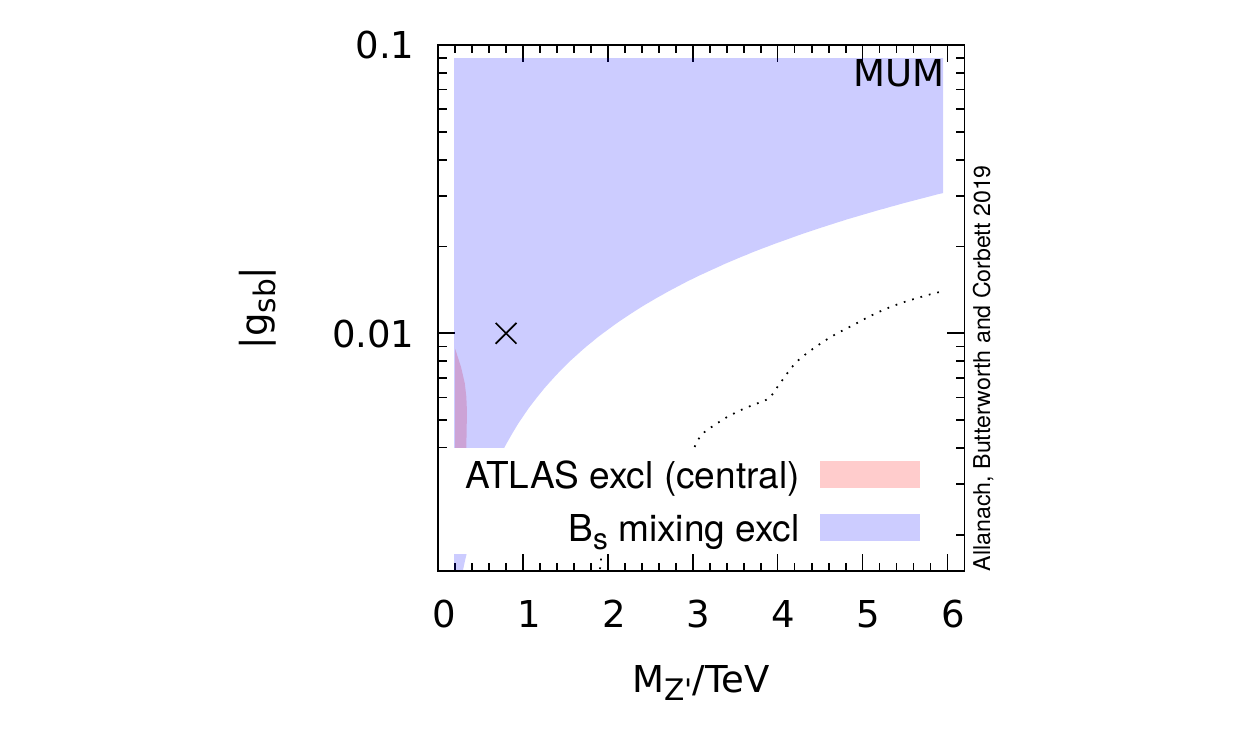}}
    \put(0.05,0.95){(a)}
    \put(0.65,0.95){(b)}
    \put(0.4,0.45){(c)}    
  \end{picture}
\end{center}
\caption{\label{fig:const} Collider constraints on the central fits of the
  models to NCBAs for (a) the MDM model, (b) the MUM model and (c) the
  TFHMeg. The allowed region is shown in white.  
  Everywhere throughout
  the parameter
  plane shown, the third parameter is fixed   by the central fit to the NCBAs: 
  $g_{\mu\mu}$ as in Eq.~\protect\ref{constraint} in (a) and (b), and
  $g_F$ as in Eq.~\ref{y3const} in (c), each with $x=1.06$.
  The region marked `non-perturbative' in (a) has
  $\Gamma_{Z^\prime}/M_{Z^\prime}>1$, whereas the dotted lines
  display where $\Gamma_{Z^\prime}/M_{Z^\prime}=0.1$ for each model (this
  quantity increases toward the right-hand side of each plot).
  The $B_s$ mixing constraint from
  Eq.~\protect\ref{bsmix} excludes the `$B_s$ mixing excl' region. In
  (c), the region excluded by the LEP lepton flavour universality is shown by the
  legend as
  `LEP LFU excl' and is calculated in Ref.~\cite{Allanach:2018lvl}.
  The ATLAS 139 fb$^{-1}$ di-muon $Z^\prime$ exclusion region is marked `ATLAS
  excl (central)' in each plot. 
  Varying the NCBA fit to be $2\sigma$ toward the SM limit results in the smaller
  `$(-2\sigma)$' exclusion region. 
The dark crosses show
  the locations of the example points listed in Table~\ref{tab:exampleSig}. In
  (c), we also display the region excluded by {\sc Contur} at the 95$\%$
  $CL_s$ level.}
\end{figure}
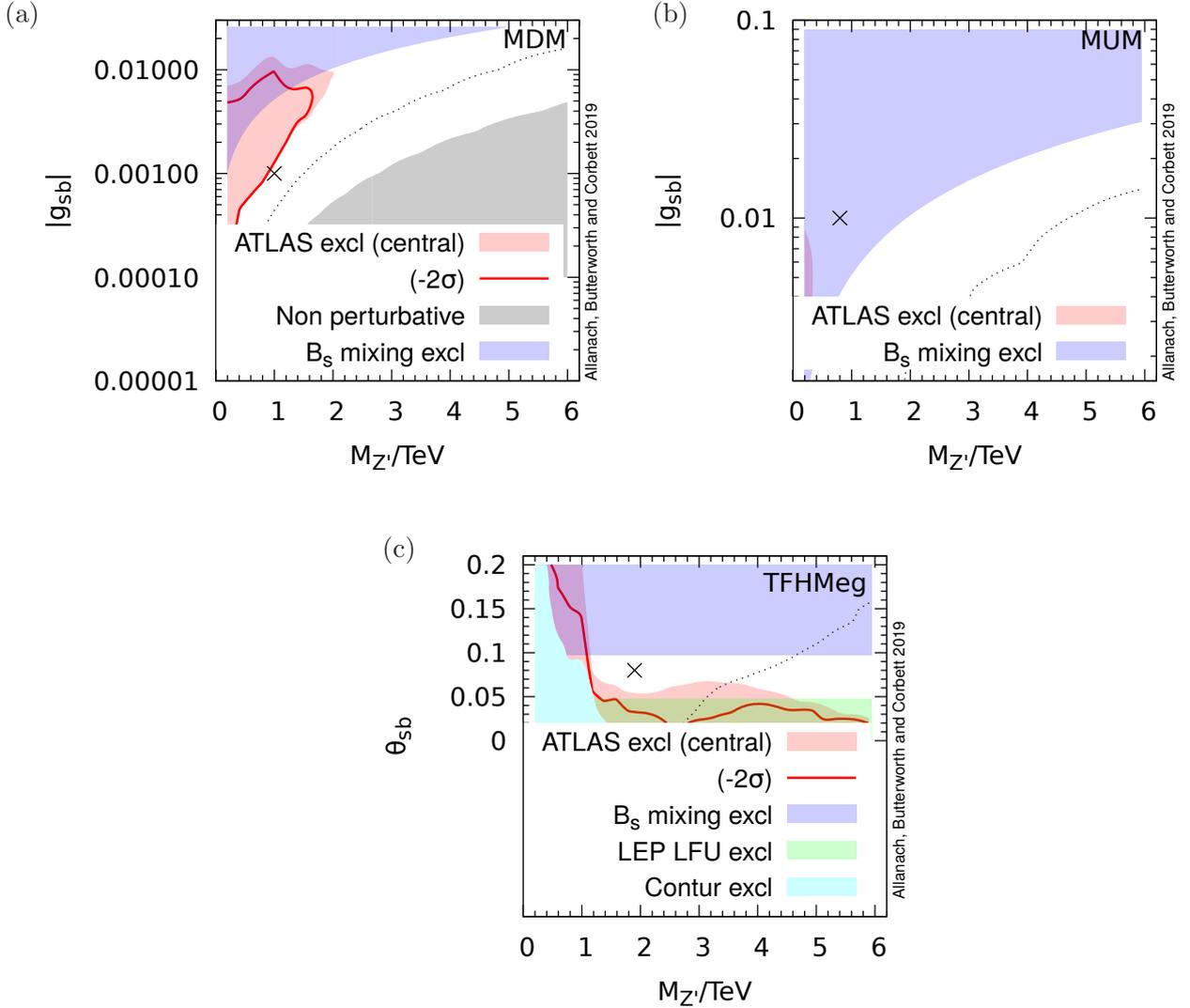
Fig.~\ref{fig:const}a displays the
collider constraints on the MDM model. 
We see that 
the bounds from the ATLAS di-muon
$Z^\prime$ search rule out a significant portion of parameter space that fits
the central value of the NCBAs and is otherwise allowed.
The shape of the various regions shown in Fig.~\ref{fig:const}a can be
understood by looking at the properties of the model across parameter space,
as shown in Fig.~\ref{fig:mdmProp}.

Since the $Z^\prime$ is produced through the quark
coupling, the higher $g_{sb}$, the higher the $Z^\prime$ production
cross-section, although it is suppressed by higher
values of $M_{Z^\prime}$ through the PDFs, as shown in 
Fig.~\ref{fig:mdmProp}c. Fitting the NCBAs means that $g_{\mu\mu}$ is small at
small $M_{Z^\prime}$ and large $|g_{sb}|$, as displayed by
Fig.~\ref{fig:mdmProp}d. This region has small $BR(Z^\prime \rightarrow \mu^+
\mu^-)$, as Fig.~\ref{fig:mdmProp}b shows, which limits the exclusion of the
ATLAS di-muon search in the top left-hand corner of
Fig.~\ref{fig:const}a. Conversely, the region with large $M_{Z^\prime}$ and
small $|g_{sb}|$ requires such a large value of $g_{\mu\mu}$ that the model
becomes non-perturbative (where the $Z^\prime$ width is equal to or larger than the mass), and we could not trust our results
there. Fortunately, this does not impact any of the bounds we have derived.

Constraints on the MUM model are summarised in Fig.~\ref{fig:const}b. We see
here that the $B_s$ mixing constraint already covers all of the region which
the ATLAS di-muon search excludes (which is hardly visible in the plot), in contrast to the MDM model shown in Fig.~\ref{fig:const}a.
The production processes do not benefit from the large $b \bar b$ contribution
present in the MDM model, as Table~\ref{tab:exampleSig} illustrates. This is
essentially because the MDM model has a $Z^\prime \bar b b$ coupling $\propto
g_{sb} / |V_{ts}|$ i.e.\ enhanced by $1/|V_{ts}| \sim 25$. 
We may understand the shape of the ATLAS constraint by referring to
Fig.~\ref{fig:mumProp} in Appendix~\ref{sec:propMUM}: the branching ratio into muons increases for smaller
$|g_{sb}|$ and larger $M_{Z^\prime}$, which competes with the cross-section
which increases toward the top left-hand corner of
Fig.~\ref{fig:const}b. Everywhere that the ATLAS di-muon constraints are active,
the $Z^\prime$ is narrow. 

Combined constraints on the TFHMeg are shown in Fig.~\ref{fig:const}c. We see
that the ATLAS di-muon search has a strong effect on the parameter space when
combined with the $B_s$ mixing constraint: $M_{Z^\prime}> 1.2$ TeV, for a
central fit to the NCBAs. The region excluded by LEP flavour universality was
calculated in Ref.~\cite{Allanach:2018lvl}, and occurs because the $Z$ picks
up small differences in its couplings to electrons as compared to muons due to
$Z-Z^\prime$ mixing. The model is non-perturbative for $M_{Z^\prime}\geq 8.4$
TeV~\cite{Allanach:2018lvl}. The white region is a relatively small portion of
parameter space, but really one should take the weaker limits at
`$(-2\sigma)$', given the possibility of statistical variations of the fit to the
NCBAs. The region to the right-hand side of the dotted line has
$\Gamma/M_{Z^\prime} > 0.1$, and so involves an extrapolation of the fit to
data given in Eq.~\ref{lims} for this region (rather than an interpolation).

{\sc Contur} exclusion limits are displayed at
the left-hand side of the figure and the excluded region is marked `Contur
excl'. There are {\em no}\/ such exclusion limits in the parameter region shown for
MDM or MUM, as the detailed {\sc Contur} plots in Fig.~\ref{fig:contur}
show. The {\sc Contur} constraints show `due diligence', in that the
interesting parameter space is not yet ruled out by a large number of LHC
SM measurements. Even though some measurements  
do receive BSM contributions to the fiducial cross section
(for example the ATLAS 13~TeV $t\bar{t}b\bar{b}$~\cite{Aaboud:2018eki} and 
8~TeV di-lepton-plus-di-jet measurements~\cite{Aaboud:2016ftt}, and the CMS
13~TeV $t\bar{t}$ measurement~\cite{Khachatryan:2016mnb}),
the measurement precision is not yet sufficient to have a strong
exclusion impact. Such exclusion as there is comes mainly from the
ATLAS 8~TeV high-mass Drell-Yan measurement~\cite{Aad:2016zzw}, and thus does not have the
reach of the 13~TeV full Run II search.

\section{Summary\label{sec:conc}}

Our focal results are the combined dominant constraints on $Z^\prime$ models
(the MUM model, the MDM model and the TFHMeg)
which fit the NBCAs and are
shown in Fig.~\ref{fig:const}.
The $B_s$ mixing constraint is important, as
well as a recent ATLAS $Z^\prime \mu^+ \mu^-$ search, performed on 139
fb$^{-1}$ of 13 TeV $pp$ LHC collisions. The ATLAS search 
is probing the otherwise allowed parameter space of the 
MDM simplified model, and we may expect the TeV HL-LHC to increase coverage 
of the parameter space~\cite{Allanach:2017bta,Allanach:2018odd}. On the other hand, the MUM
simplified model is currently more constrained by the $B_s$ mixing constraint,
and likely will require an increase in energy~\cite{Allanach:2017bta,Allanach:2018odd} (for example to HE-LHC~\cite{CidVidal:2018eel} or
FCC,\cite{Mangano:2018mur}) for di-muon searches to probe the remaining
parameter space. 
The $B_s$ mixing constraint is particularly constraining, but there has been
significant movement on it in the last four years, mainly due to different
estimates of the SM contribution. The 95$\%$ $CL$ bound has been
$M_{Z^\prime}/g_{sb}>148, 600, 194$ TeV, respectively. 
We might therefore expect further movement upon the bound in the future, and
this could 
have a large impact on the constraints. Taking the current bound of 194 TeV at
face value, we extract (from the `(-2$\sigma$)' bounds in
Fig.~\ref{fig:const}c and from Fig.~\ref{fig:tfhmProp}b) that $\sigma \times
BR(Z^\prime \rightarrow \mu^+ \mu^-)\geq 2.6 \times
10^{-3}$ fb in the TFHMeg at a centre of mass energy $\sqrt{s}=13$ TeV. The
lower bound is saturated for $M_{Z^\prime}=3.5$ TeV, $\theta_{sb}=0.1$. 
At
$\sqrt{s}=14$ TeV, we estimate  (from this point)
the minimum cross-sections
in Table~\ref{tab:ex}. Since the nominal integrated luminosity for the HL-LHC
is ${\mathcal L}=3000$ fb$^{-1}$, we may expect at least $S=30$ signal
$Z^\prime \rightarrow \mu^+ \mu^-$ 
events. 
We are therefore hopeful of 
the TFHMeg HL-LHC $Z^\prime$ search prospects\footnote{Background estimates,
  which are beyond the scope of this work,
would be required to properly calculate the sensitivity.}.
\begin{table}\begin{center}
  \begin{tabular}{|c|ccccc|} \hline
    Channel & $\mu^+ \mu^-$ & $t \bar t$ & $\tau^+ \tau^-$ & $b \bar b$ &
    $\bar \nu_i \nu_j$\\ \hline
    $\sigma \times BR$/fb& 0.02 & 0.08 & 0.07 & 0.03 & 0.02 \\ \hline
  \end{tabular}
  \caption{\label{tab:ex} HL-LHC ($\sqrt{s}=14$ TeV) minimum fiducial cross section times 
    branching ratios for different final-state channels in the TFHMeg.} \end{center}
  \end{table}

\section*{Acknowledgements}
This work has been partially supported by STFC consolidated grants
ST/P000681/1 and ST/N000285/1. 
BCA and JMB thank the {\em (Re)interpreting the results of new physics searches at the
  LHC}\/ workshop, where part of this work was carried out.
BCA thanks A Lenz and the Cambridge SUSY Working Group for
discussions. 
This work has received funding from the European Union's 
Horizon 2020 research and innovation programme as part of the 
Marie Skłodowska-Curie Innovative Training Network MCnetITN3 (grant agreement no. 722104).

\appendix
\section{Field Definitions \label{sec:fdef}}
We use the following field definitions in terms of representations of
$SU(3)\times SU(2)_L \times U(1)_Y$: 
${\bm Q_L}=(3,\ 2,\ +1/6)=({\bm u_L}, {\bm d_L})^T$,
${\bm L_L}=(1,\ 2,\ -1/2)=({\bm \nu_L}, {\bm e_L})^T$, 
$Z^\prime=(1\ ,1\ ,0)$, where bold face
denotes a 3-dimensional vector in family space. These fields are implicitly written in the mass eigenbasis.

\section{Properties of the Models\label{sec:propMUM}}
\begin{figure}
\begin{center}
  \unitlength=\textwidth
  \includegraphics[width=0.9 \textwidth]{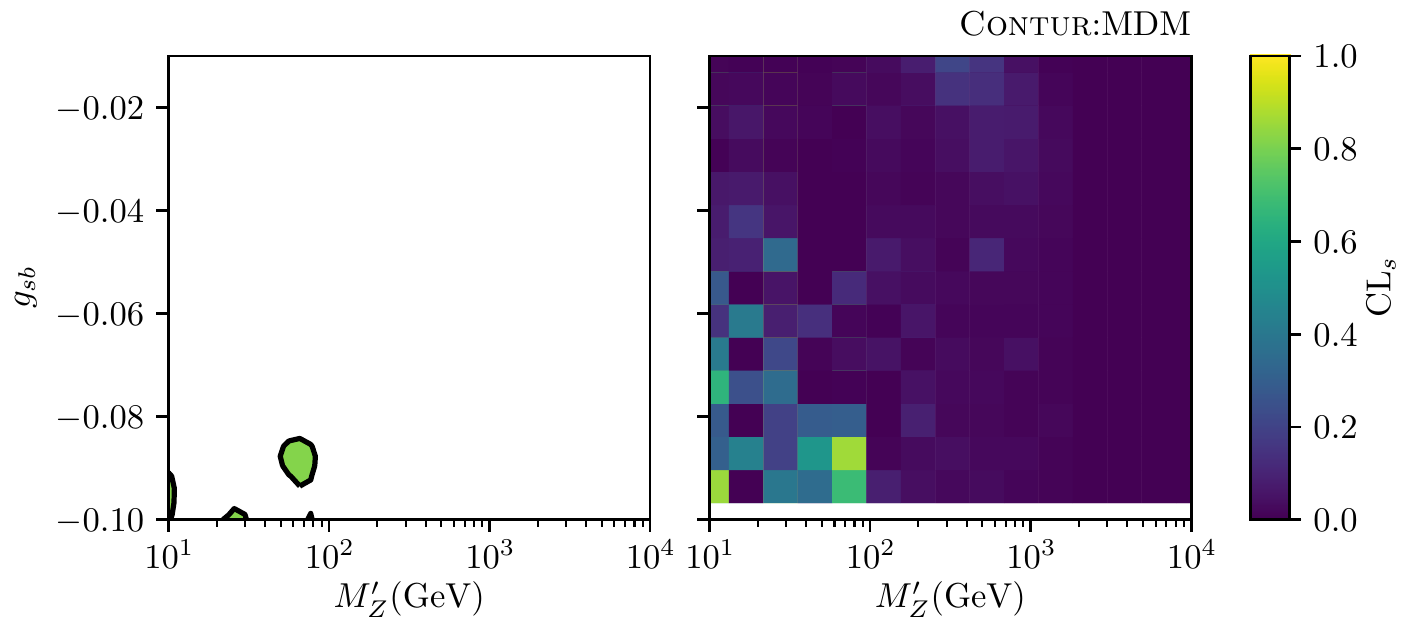}
  \includegraphics[width=0.9 \textwidth]{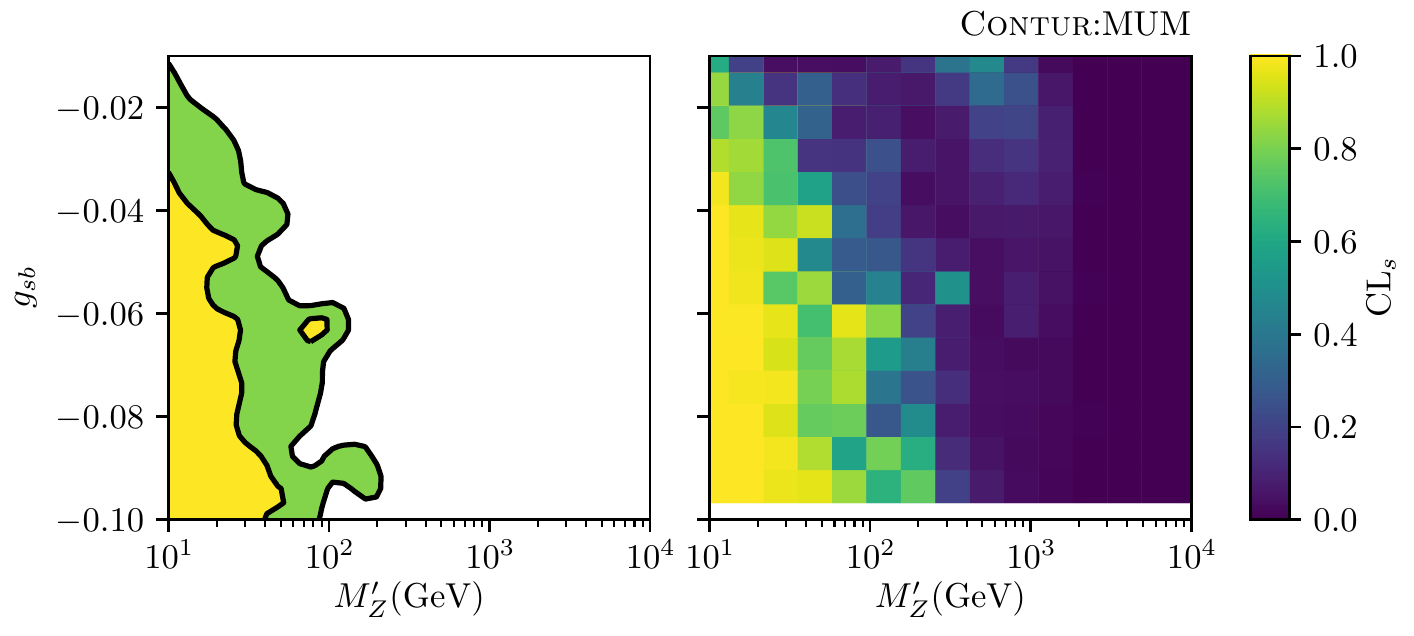}
  \includegraphics[width=0.9 \textwidth]{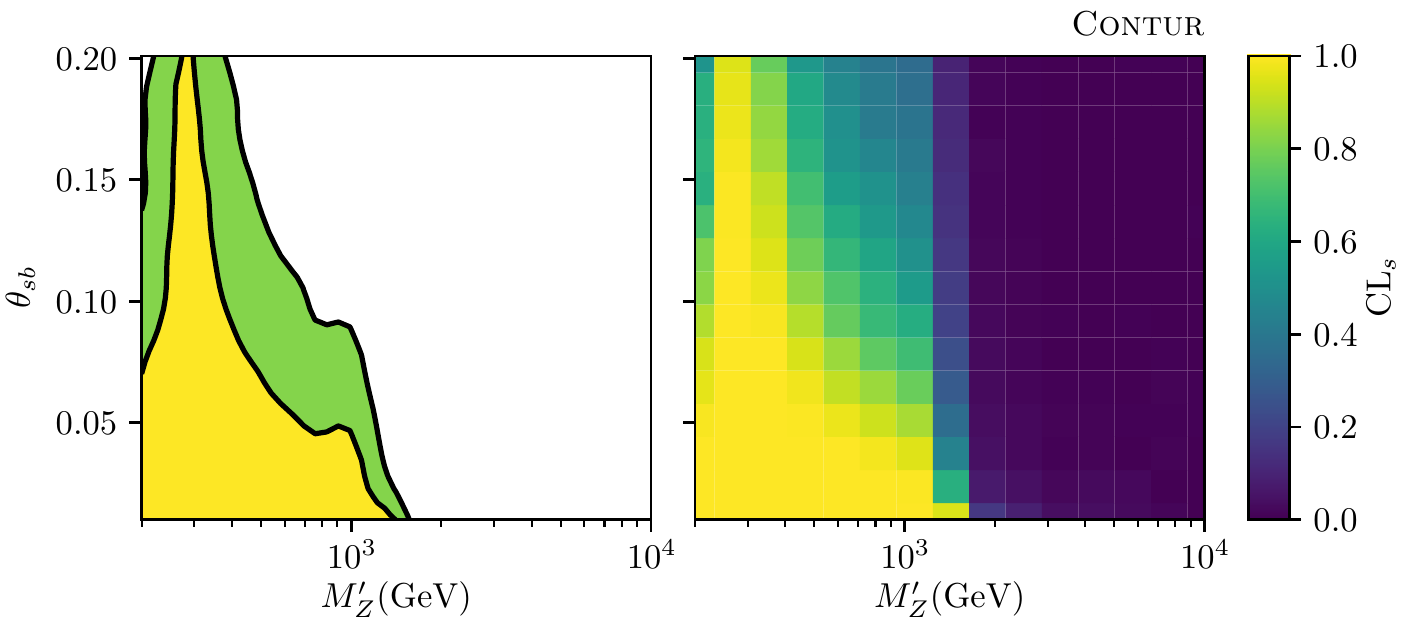}  
\end{center}
\caption{\label{fig:contur} {\sc Contur} constraints on the MDM model (top), the
  MUM model
  (middle) and the TFHMeg (bottom).
  In the left-hand panels, we show regions excluded at the $68\%$ and
  $95\%$ $CL_s$ levels. 
  In the right-hand panels, we show the $CL_s$ values, where 95$\%$
  excluded values are $>0.95$. Everywhere throughout
  the parameter
  plane shown, the NCBAs are fit by fixing $g_{\mu\mu}$ and $g_F$ as in  
  Eqs.~\ref{constraint},\ref{y3const}, respectively, for $x=1.06$}
\end{figure}
We display the {\sc Contur} constraints on the different models in
Fig.~\ref{fig:contur}. There are essentially no constraints upon the MDM
model, whereas the MUM model is somewhat constrained for $M_{Z^\prime}<100$
GeV. The strongest constraints are upon the TFHMeg, which extend to
$M_{Z^\prime}=1.5$ TeV, for low $\theta_{sb}$ (where $g_F$ is high). 

\begin{figure}
\begin{center}
\unitlength=\textwidth
\begin{picture}(1,1)(0.05,0.13)
\put(-0.05,0.5){\includegraphics[width=0.65 \textwidth]{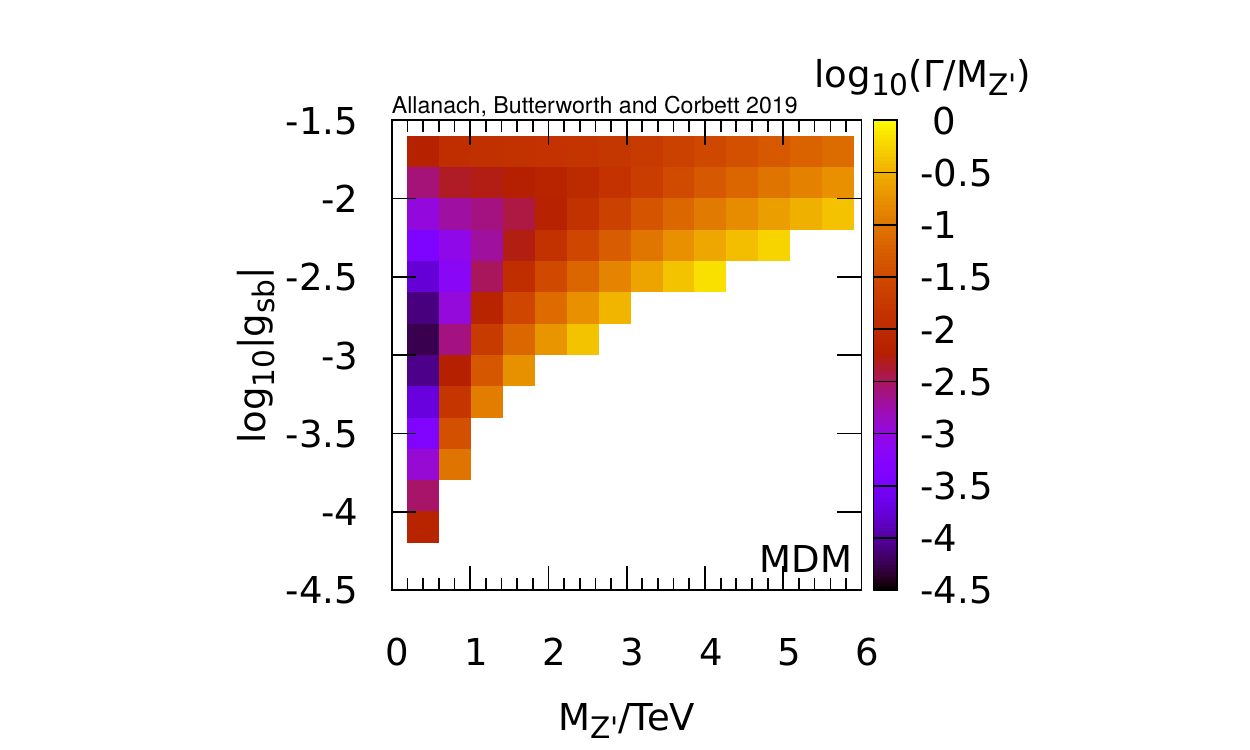}}
\put(0.5,0.5){\includegraphics[width=0.65 \textwidth]{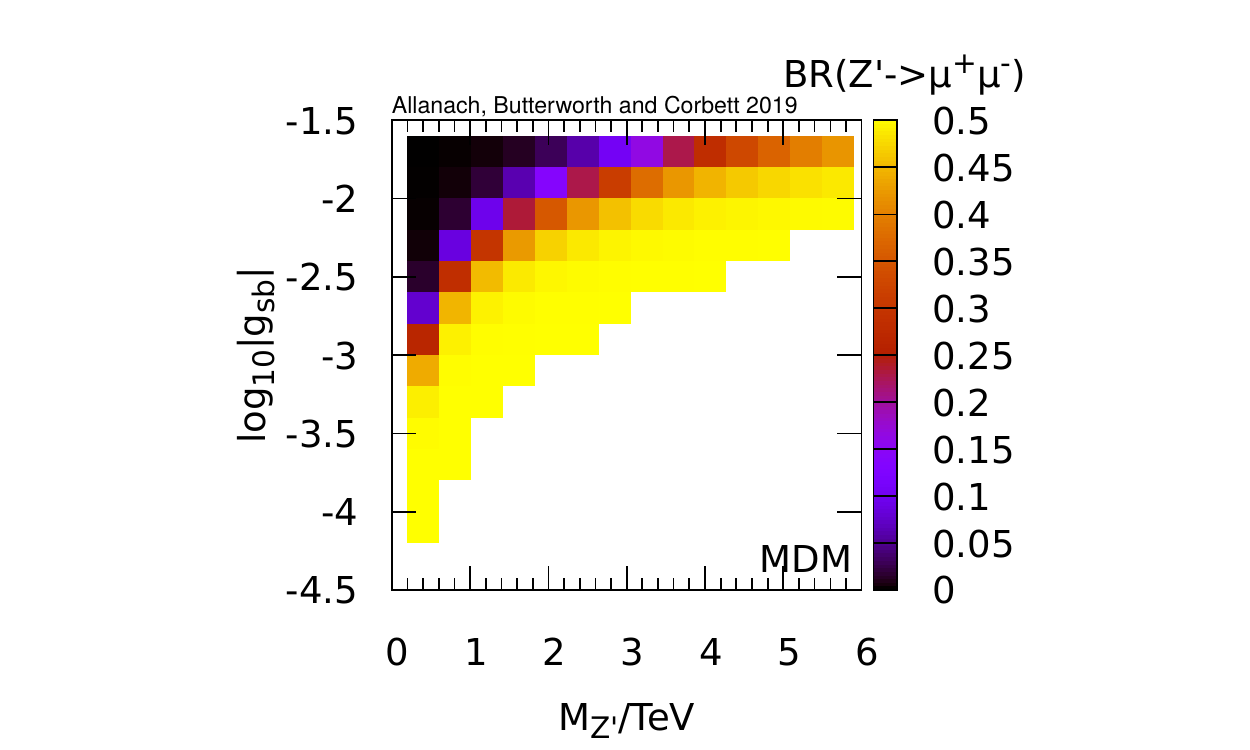}}
\put(-0.05,0.13){\includegraphics[width=0.65 \textwidth]{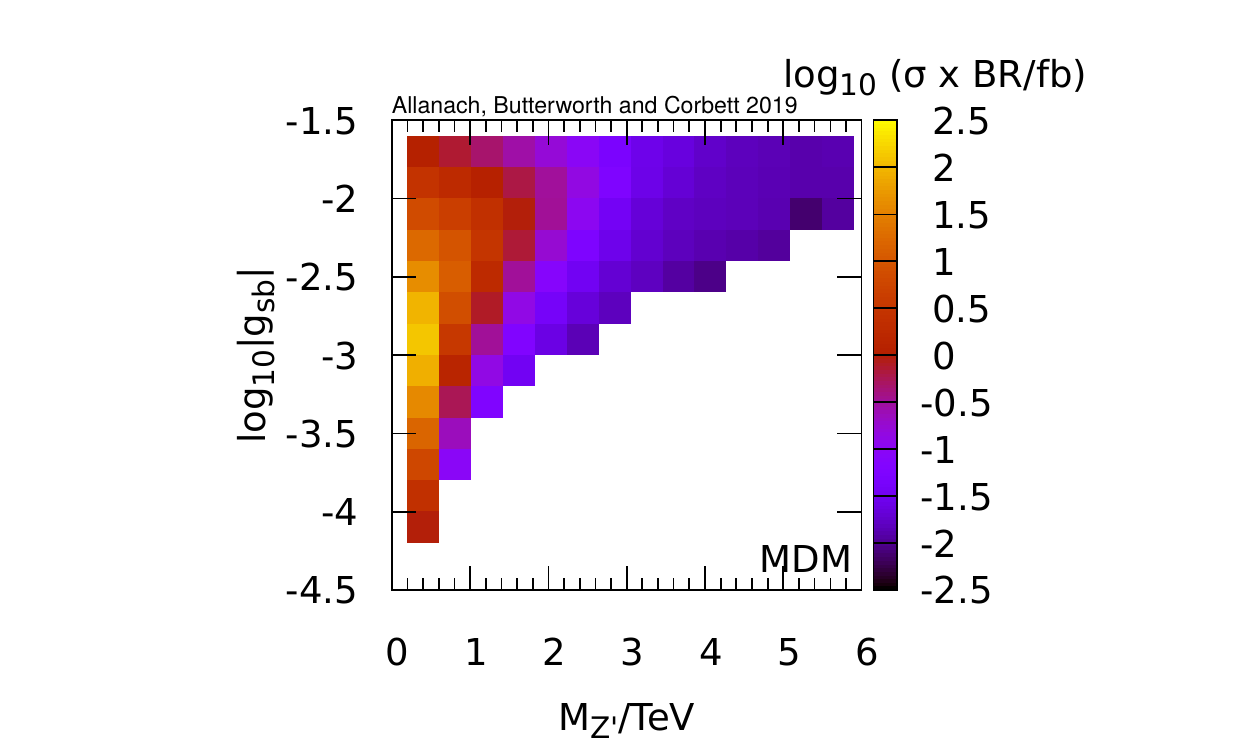}}
\put(0.5,0.13){\includegraphics[width=0.65 \textwidth]{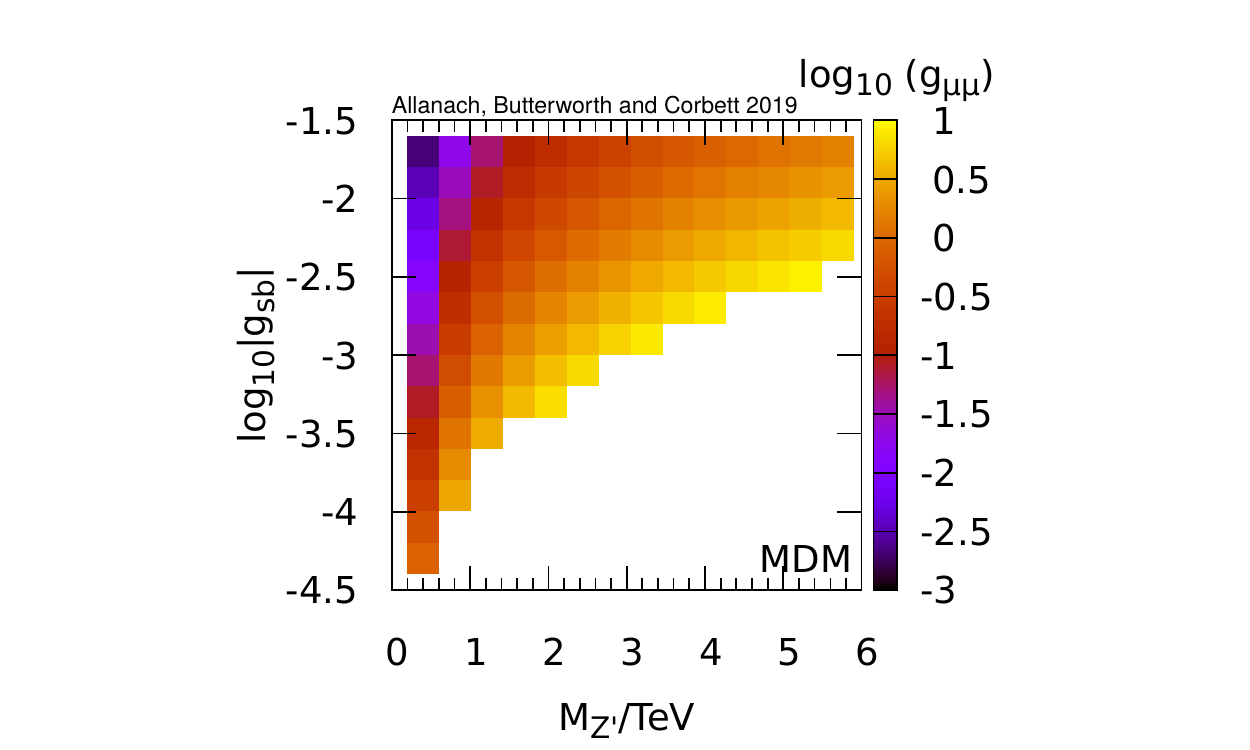}}
\put(0.05,0.85){(a)}
\put(0.55,0.85){(b)}
\put(0.05,0.48){(c)}
\put(0.55,0.48){(d)}
\end{picture}
\end{center}
\caption{Properties of the central fit of the MDM model to NCBAs. In (a), we show the $Z^\prime$ width divided by
  its mass, $\Gamma/M_{Z^\prime}$. In (b), the branching ratio into di-muons
  is shown, in (c) the fiducial $Z^\prime$ production cross section multiplied
  by its branching ratio into di-muons is displayed. (d) shows 
  $g_{\mu\mu}$ coming from the central fit to NCBAs. The white region is
  non-perturbative. 
  \label{fig:mdmProp}}
\end{figure}

\begin{figure}
\begin{center}
\unitlength=\textwidth
\begin{picture}(1,0.85)(0,0.13)
\put(-0.05,0.5){\includegraphics[width=0.65 \textwidth]{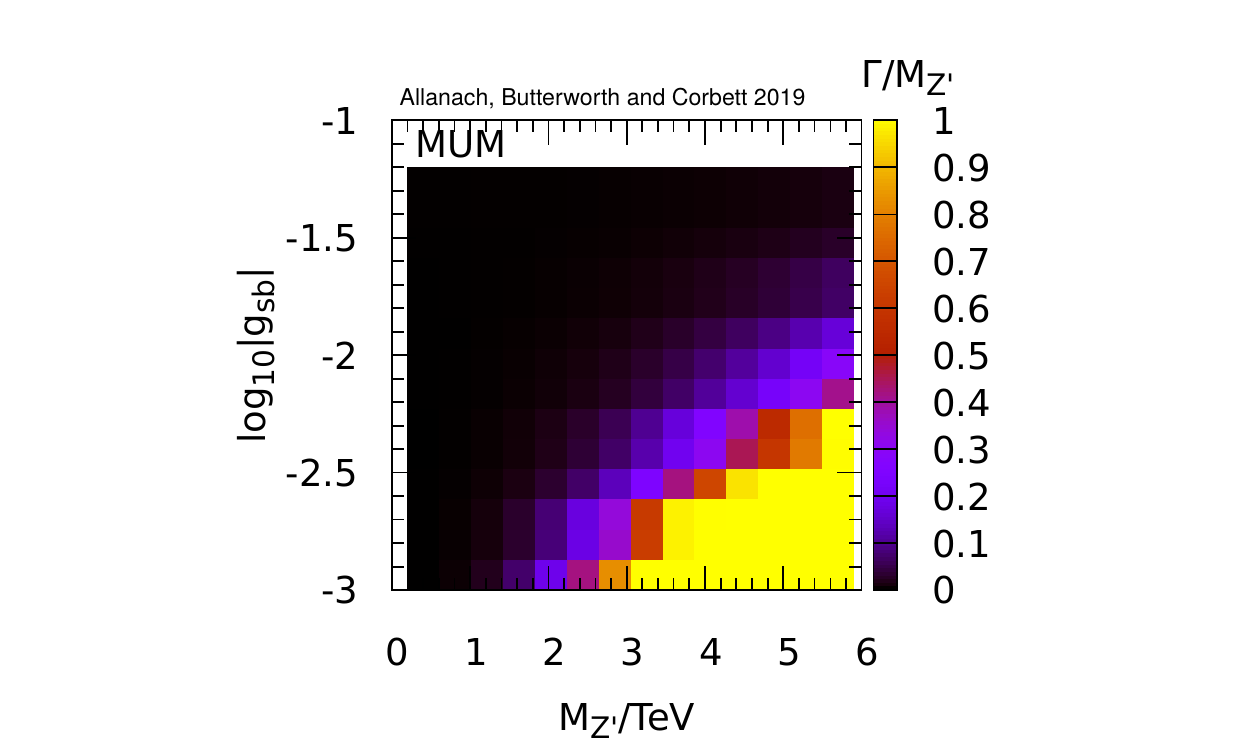}}
\put(0.5,0.5){\includegraphics[width=0.68 \textwidth]{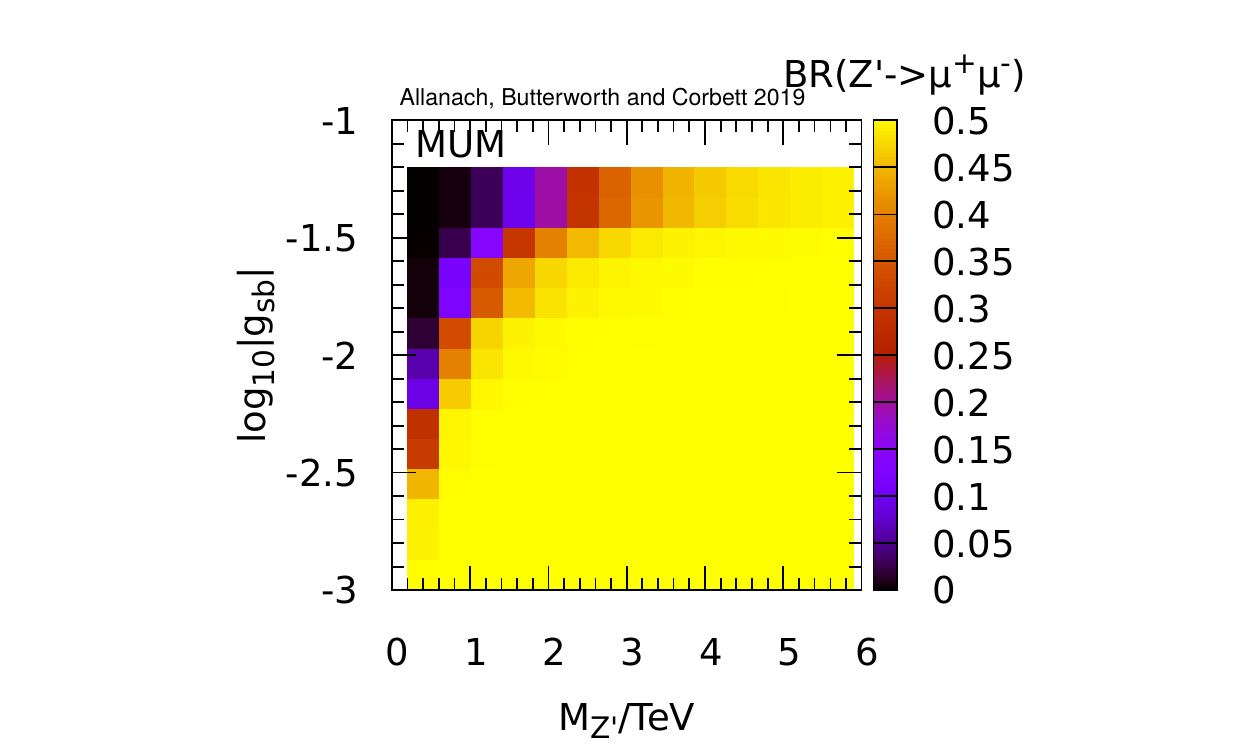}}
\put(-0.05,0.13){\includegraphics[width=0.65 \textwidth]{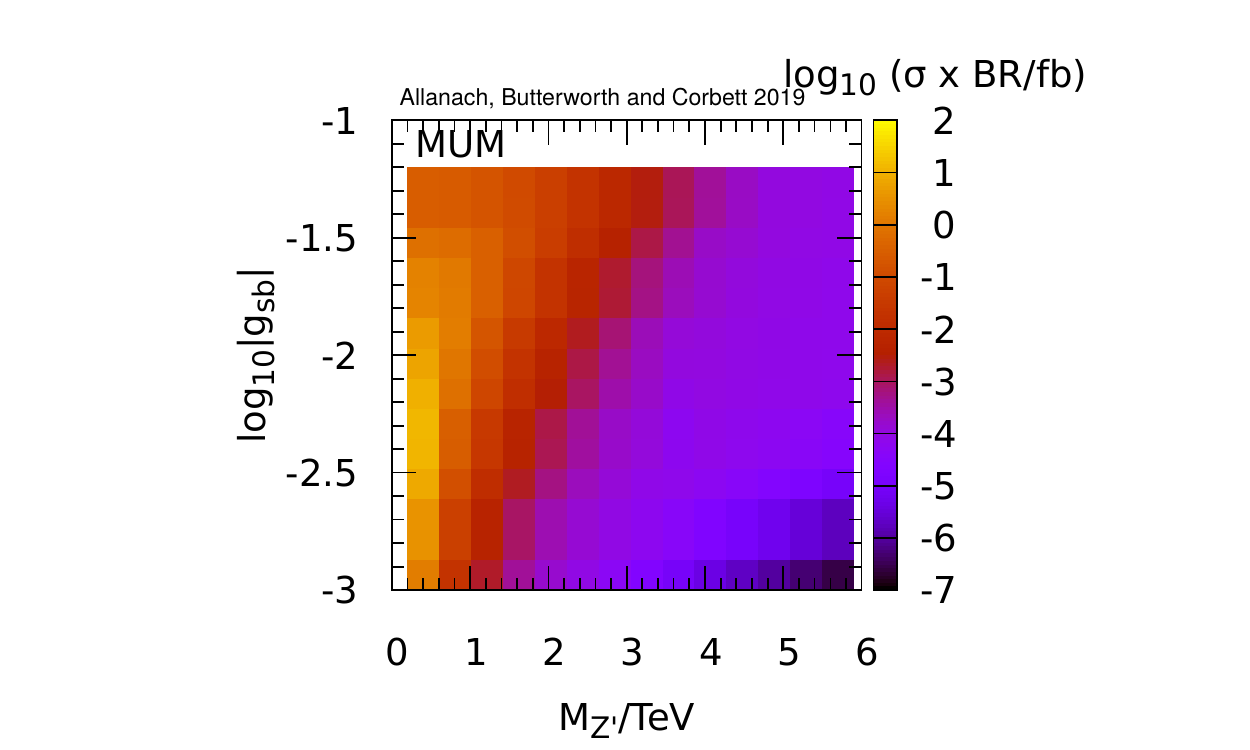}}
\put(0.5,0.13){\includegraphics[width=0.65 \textwidth]{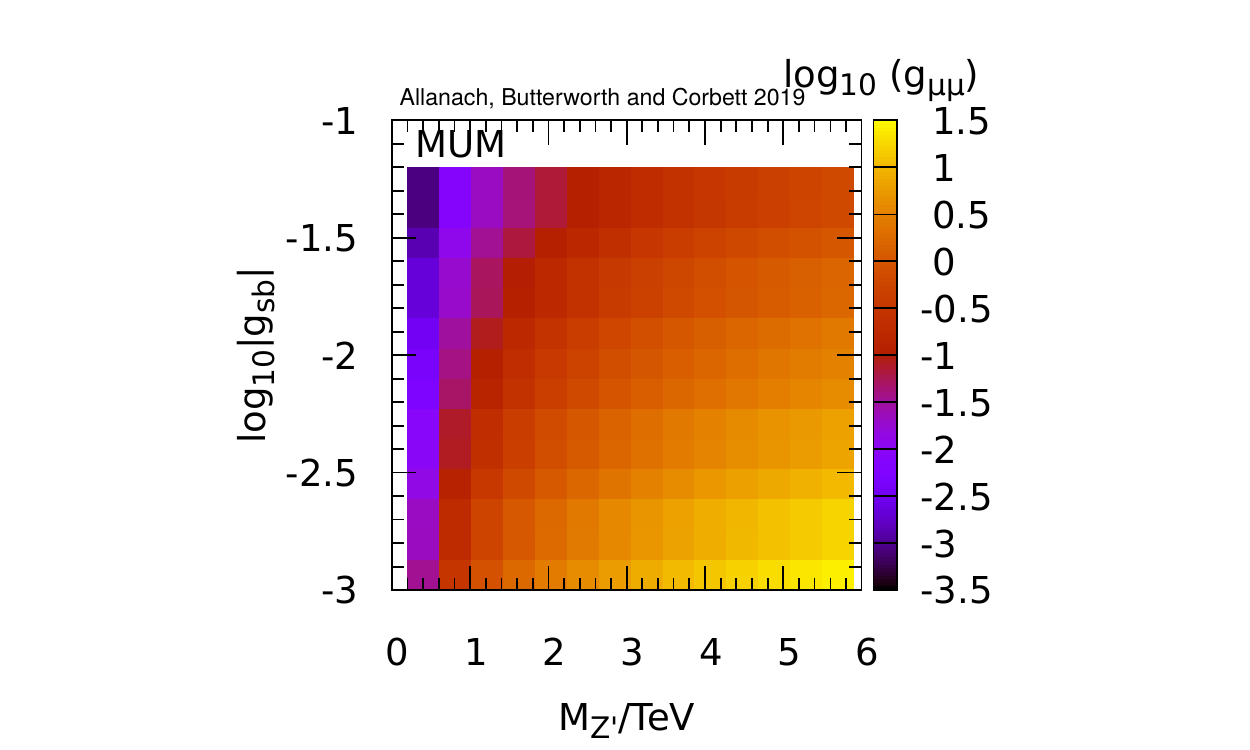}}
\put(0.05,0.85){(a)}
\put(0.55,0.85){(b)}
\put(0.05,0.48){(c)}
\put(0.55,0.48){(d)}
\end{picture}
\end{center}
\caption{\label{fig:mumProp}
Properties of the central fit of the MUM model to NCBAs. In (a), we show the $Z^\prime$ width divided by
  its mass, $\Gamma/M_{Z^\prime}$. In (b), the branching ratio into di-muons
  is shown, in (c) the fiducial $Z^\prime$ production cross section multiplied
  by its branching ratio into di-muons is displayed. (d) shows 
  $g_{\mu\mu}$ coming from the central fit to NCBAs.}
\end{figure}
\begin{figure}
\begin{center}
\unitlength=\textwidth
\begin{picture}(1,0.85)(0,0.13)
\put(-0.05,0.5){\includegraphics[width=0.65 \textwidth]{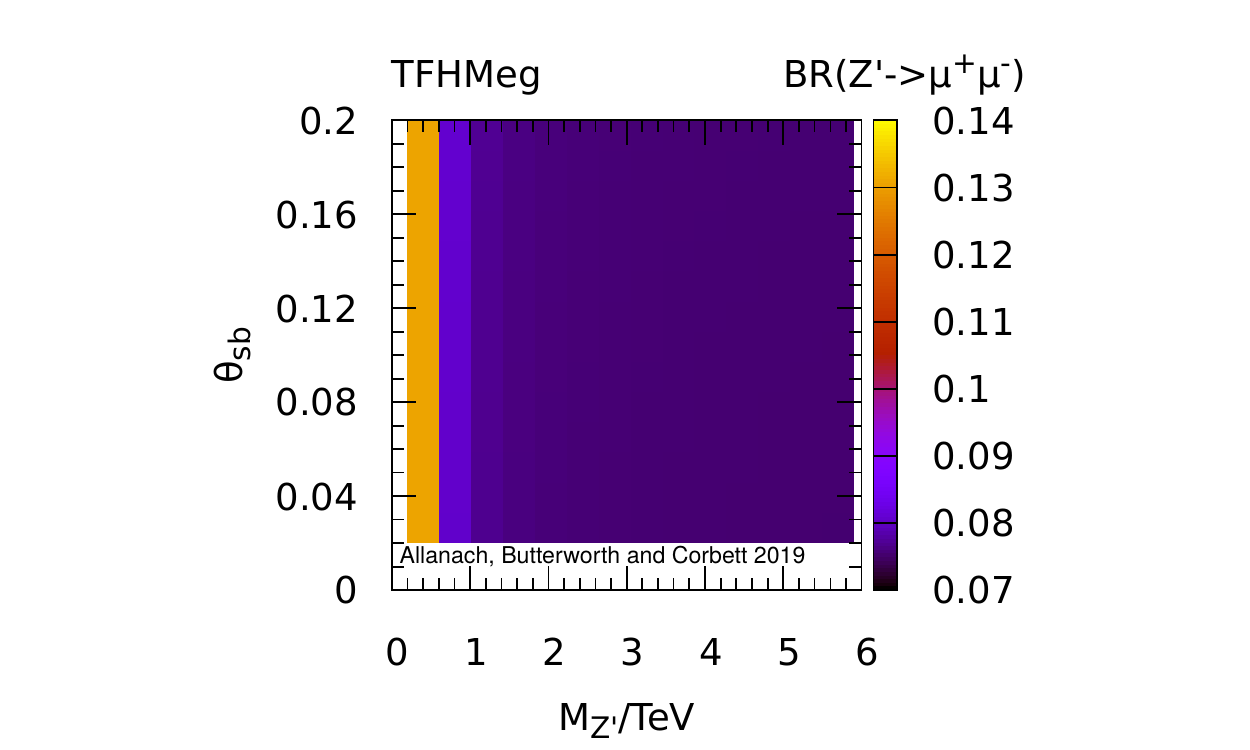}}
\put(0.5,0.5){\includegraphics[width=0.68 \textwidth]{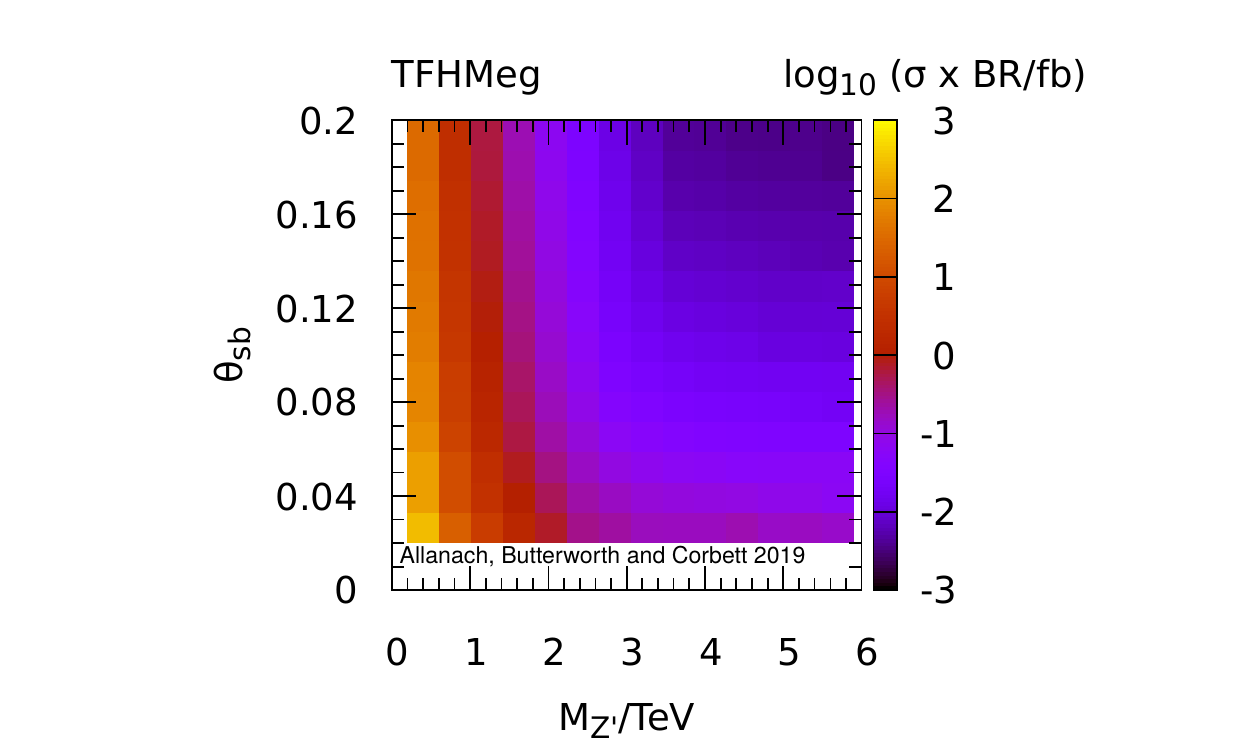}}
\put(-0.05,0.13){\includegraphics[width=0.65 \textwidth]{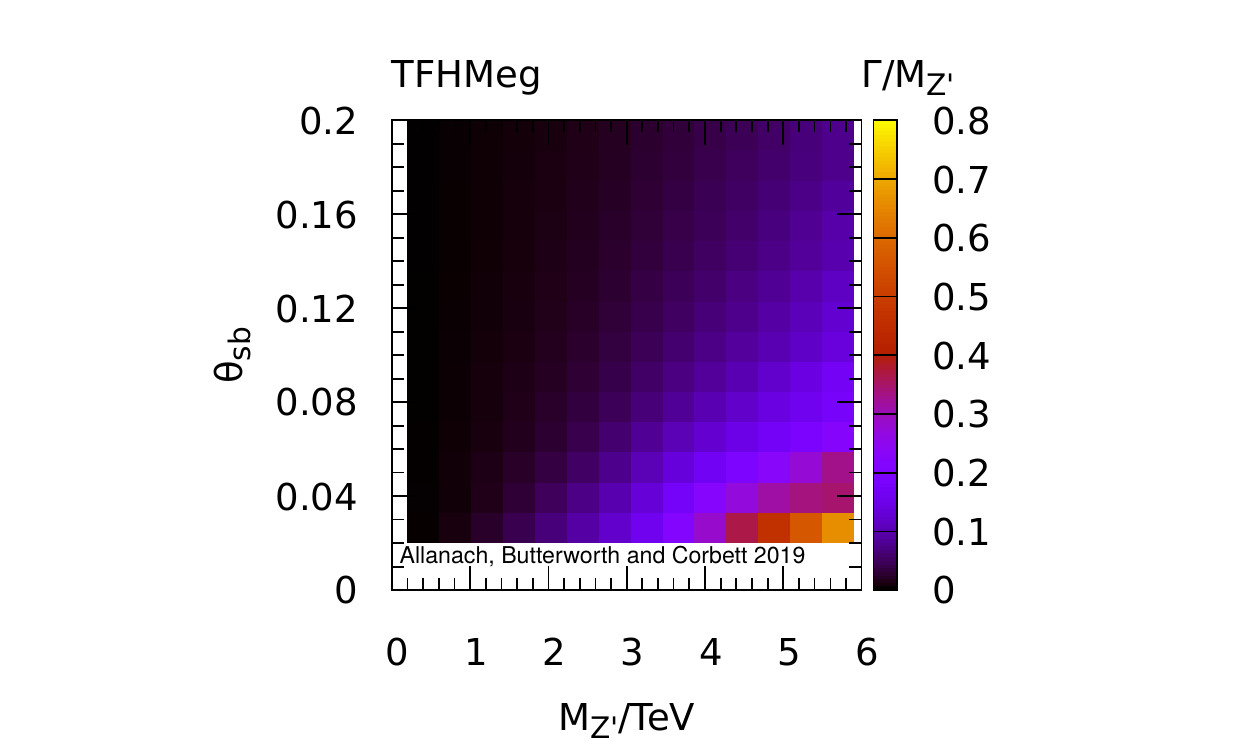}}
\put(0.5,0.13){\includegraphics[width=0.65 \textwidth]{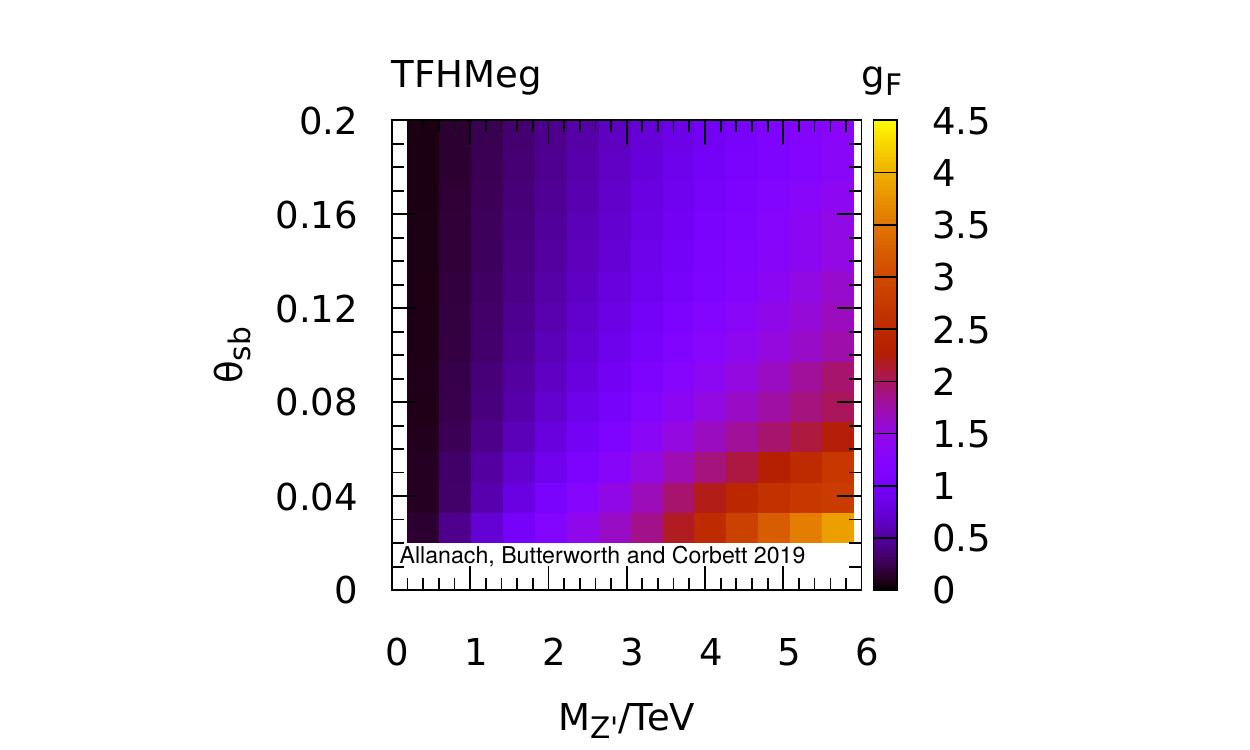}}
\put(0.05,0.85){(a)}
\put(0.55,0.85){(b)}
\put(0.05,0.48){(c)}
\put(0.55,0.48){(d)}
\end{picture}
\end{center}
\caption{Properties of the TFHMeg. Everywhere throughout
  the parameter
  plane shown, the NCBAs are fit by fixing $g_F$ as in 
  Eq.~\ref{y3const} for $x=1.06$ and shown in
  panel (d). \label{fig:tfhmProp}}
\end{figure}

We display some properties of the MDM model across parameter space in
Fig.~\ref{fig:mdmProp}, some of the MUM model in Fig.~\ref{fig:mumProp} and
some of the TFHMeg in Fig.~\ref{fig:tfhmProp}.

\bibliographystyle{JHEP-2}
\bibliography{currentLim}

\end{document}